\documentclass[twocolumn]{article}

\usepackage[a4paper,left=1cm,right=1cm,top=3cm,bottom=3cm]{geometry}
\usepackage{enumitem}
\usepackage{amsmath, amssymb, amsthm}
\usepackage{graphicx}
\usepackage{faktor}
\usepackage{multirow}
\usepackage{hyperref}
\usepackage{url}
\usepackage{booktabs}
\usepackage{subfig}
\usepackage{color}
\newcommand{\etal}{\textit{et al.}}
\newcommand{\afv}{A\(^\ast\)FV}
\newcommand{\calR}{\mathcal{R}}
\newcommand{\bfc}{\mathbf{c}}
\newcommand{\relu}{\mathsf{ReLU}}

\theoremstyle{definition}
\newtheorem{exmp}{Example}[section]

\usepackage{color}
\newif\ifnotes

\ifnotes
\newcommand{\caesar}[1]{{\color{blue}{#1}}}
\else
\newcommand{\caesar}[1]{#1}
\fi

\begin{document}
%
\title{Towards the AlexNet Moment for Homomorphic Encryption: HCNN, the First Homomorphic CNN on Encrypted Data with GPUs}
%
%
%


\author{
    Ahmad~Al~Badawi,
	Chao~Jin,
	Jie~Lin,
	Chan~Fook~Mun,
	Sim~Jun~Jie, \\
	Benjamin~Hong~Meng~Tan,
	Xiao~Nan,
	Khin~Mi~Mi~Aung, \\	and~Vijay~Ramaseshan~Chandrasekhar
	\thanks{V. R. Chandrasekhar and X. Nan contributed to this work while at I$^2$R. Manuscript revised \today.}}
\markboth{IEEE TRANSACTIONS ON EMERGING TOPICS IN COMPUTING}{AL BADAWI  \MakeLowercase{\textit{et al.}}: Towards the AlexNet Moment for Homomorphic Encryption: HCNN, the First Homomorphic CNN on Encrypted Data with GPUs}

\maketitle

\maketitle

\begin{abstract}
Deep Learning as a Service (DLaaS) stands as a promising solution for cloud-based inference applications. In this setting, the cloud has a pre-learned model whereas the user has samples on which she wants to run the model. The biggest concern with DLaaS is the user privacy if the input samples are sensitive data. We provide here an efficient privacy-preserving system by employing high-end technologies such as Fully Homomorphic Encryption (FHE), Convolutional Neural Networks (CNNs) and Graphics Processing Units (GPUs).\\
FHE, with its widely-known feature of computing on encrypted data, empowers a wide range of privacy-concerned applications. This comes at high cost as it requires enormous computing power. In this paper, we show how to accelerate the performance of running CNNs on encrypted data with GPUs. We evaluated two CNNs to classify homomorphically the MNIST and CIFAR-10 datasets. Our solution achieved sufficient security level ($> 80$ bit) and reasonable classification accuracy (99\%) and (77.55\%) for MNIST and CIFAR-10, respectively. In terms of latency, we could classify an image in 5.16 seconds and 304.43 seconds for MNIST and CIFAR-10, respectively. Our system can also classify a batch of images ($>$ 8,000) without extra overhead.
\end{abstract}


%
%

\section{Introduction}\label{section:Introduction}
%
%
%
%
Deep Learning (DL) has empowered a wide range of applications from labelling systems, web searches, content filtering to recommendation systems on entertainment and e-commerce systems. The prominent feature of this technology is its ability to perform tasks that we humans can do seamlessly such as labelling, identification and recommendation. To do that, this technology relies on complex models that can capture specific features from input data. Building these models is a complex task that requires substantial domain expertise of several disciplines starting from neurology to computer science. Therefore, there is an evolving trend toward leasing instead of building these models, which made Deep Learning as a Service (DLaaS) an indispensable solution. 

The cloud stands as a good platform to either create or host pre-learned models as it offers cheap data storage, near-zero deployment cost and high computational services~\cite{jadeja2012cloud}. However, it has some drawbacks and raises important questions that need to be resolved. One of the main concerns is that cloud platforms do not guarantee data privacy. In DLaaS setting, the user uploads her data to the cloud which in turn evaluates the model on the input data and sends the results back to the user. At any step along this process, there are numerous opportunities for attackers to compromise the data.


The main objective of our work is to provide \caesar{a} secure, efficient and non-interactive solution to the aforementioned problem. We use Fully Homomorphic Encryption (FHE)~\cite{STOC:Gentry09} - which is a new cryptographic technique that allows one to evaluate functions on encrypted data without decryption or access to the secret key - to evaluate Convolutional Neural Networks (CNNs) on encrypted data. Our solution reveals nothing about the input, say $x$, to the cloud nor it does about the model, say $f$, to the client except what can be learned from input $x$ and $f(x)$. Our solution is efficient and reasonably practical especially if multiple predictions are to be made simultaneously. Lastly, it can be considered non-interactive as the client needs only to interact with the cloud to provide the input data and receive the output.

FHE-based privacy-preserving DLaaS was considered previously by Graepel \etal~\cite{ICISC:GraLauNae12} and Aslett \etal~\cite{ARXIV:AslEspHol15}. Following them, Dowlin \etal~\cite{MSFT:DGL+16} proposed CryptoNets, the first CNN over encrypted images, providing a method to do the inference phase of DL. The main drawback of these FHE-based solutions is the computational overhead. For instance, CryptoNets required 570 seconds to evaluate a FHE-friendly model on encrypted samples from the MNIST dataset~\cite{MNIST} at security level (80-bit)\caesar{\footnote{Note that 80-bit security level is widely accepted in the literature due to performance constraints. However, it is considered a legacy security parameter in conventional cryptography by NIST and other standardization bodies.}}. To this end, we design a solution that overcomes the aforementioned problems. For instance, our solution is more efficient and requires only 5.16 seconds to evaluate the MNIST model with security level $>$ 80 bits. Inspired by Krizhevsky's AlexNet \etal~\cite{NIPS:KriSutHin12} who showed how image classification is viable by running CNN on GPUs, our main contribution in this work is to show that privacy-preserving DL is not only possible on GPUs but can also be dramatically accelerated. 

\subsection{Our Contributions}
\begin{enumerate}\itemsep=0.1mm
    \item We present the first GPU-accelerated HCNN that runs a pre-learned model on encrypted images.
    \item We combine a rich set of optimization techniques such as quantized NNs with low-precision training, optimized choice of FHE scheme and parameters, and a GPU-accelerated implementation.
    \item We present 2 new FHE-friendly CNNs to classify encrypted images from MNIST and CIFAR-10 datasets. Our MNIST HCNN is only 5 layers deep for both training and inference with 43-bit plaintext modulus, smaller than CryptoNets~\cite{MSFT:DGL+16}, which uses 9 layers during training with 80-bit plaintext modulus. For CIFAR-10, we provide a novel 11-layer network.
    \item Experiments show that our MNIST HCNN can evaluate an image in 5.16 seconds with 99\% accuracy. Our CIFAR-10 HCNN requires 304.43 seconds and offers 77.55\% accuracy.
\end{enumerate}

\subsection{Related Work}
The research in the area of privacy-preserving DL can be roughly divided into two camps: those using homomorphic encryption or combining it with secure MPC techniques.
Most closely related to our work are CryptoNets by Dowlin~\etal~\cite{MSFT:DGL+16}, FHE-DiNN by Bourse \etal~\cite{C:BMMP18} and E2DM by Jiang \etal~\cite{Jiang:2018:SOM:3243734.3243837}, who focus on using only fully homomorphic encryption to address this problem.
Dowlin \etal~\cite{MSFT:DGL+16} were the first to propose using FHE to achieve privacy-preserving DL, offering a framework to design NNs that can be run on encrypted data.
They proposed using polynomial approximations of the most widespread \(\relu\) activation function and using pooling layers only during the training phase to reduce the circuit depth of their neural network.
However, they used the YASHE\(^{\prime}\) scheme by Bos \etal~\cite{IMA:BLLN13}, which is no longer secure due to attacks proposed by Albrecht \etal~\cite{C:AlbBaiDuc16}.
Also, they require a large plaintext modulus of over 80 bits to accommodate the output result of their neural network's.
This makes it very difficult to scale to deeper networks since intermediate layers in those networks will quickly reach several hundred bits with such settings.

Following them, Bourse \etal~\cite{C:BMMP18} proposed a new type of NNs called Discretized Neural Networks~(DiNN) for inference over encrypted data.
Weights and inputs of traditional CNNs are discretized into elements in \(\{-1, 1\}\) and the fast bootstrapping of the TFHE scheme proposed by Chilotti~\etal~\cite{AC:CGGI16} was exploited to double as an activation function for neurons.
Each neuron computes a weighted sum of its inputs and the activation function is the sign function, \(\mathsf{sign}(z)\) which outputs the sign of the input \(z\).
Although this method can be applied to arbitrarily deep networks, it suffers from lower accuracy, achieving only \(96.35\%\) accuracy on the MNIST dataset with lower amortized performance.
Very recently, Jiang \etal~\cite{Jiang:2018:SOM:3243734.3243837} proposed a new method for matrix multiplication with FHE and evaluated a neural network on the MNIST data set using this technique.
They also considered packing an entire image into a single ciphertext compared to the approach of Dowlin~\etal~\cite{MSFT:DGL+16}, who put only one pixel per ciphertext but evaluated large batches of images at a time.
They achieved good performance, evaluating \(64\) images in slightly under \(29\) seconds but with worse amortized performance.

Some of the main limitations of pure FHE-based solutions is the need to approximate non-polynomial activation functions and high computation time.
Addressing these problems, Liu~\etal~\cite{CCS:LJLA17} proposed MiniONN, a paradigm shift in securely evaluating NNs.
They take commonly used protocols in DL and transform them into oblivious protocols.
With MPC, they could evaluate NNs without changing the training phase, preserving accuracy since there is no approximation needed for activation functions.
However, MPC comes with its own set of drawbacks.
In this setting, each computation requires communication between the data owner and model owner, thus resulting in high bandwidth usage.
In a similar vein, Juvekar \etal~\cite{USENIX:JuvVaiCha18} designed GAZELLE.
Instead of applying levelled FHE, they alternate between an additive homomorphic encryption scheme for convolution-type layers and garbled circuits for activation and pooling layers.
This way, communication complexity is reduced compared to MiniONN but unfortunately is still significant.

\caesar{Following the footsteps of GAZELLE, Pratyush et al. proposed DELPHI as a system for cryptographic inference service~\cite{USENIX:srinivasandelphi19}. DELPHI reports 22$\times$ and 9$\times$ improvement in inference latency and communication overhead on ResNet-32. DELPHI achieves this by replacing the homomorphic encryption part in GAZELLE with secret sharing. Hence, instead of transferring and computing on ciphertexts, secret shares of the model and the client's inputs are exchanged. Moreover, instead of evaluating \(\relu\) as a garbled circuit, DELPHI uses polynomial approximations of \(\relu\).}

\caesar{Another recent MPC-based oblivious inference framework, called XONN, was proposed in~\cite{riazi2019xonn}. XONN uses binarized neural networks, i.e., networks with binary weights and binary inputs. By doing so, the framework avoids expensive multiplications and uses bit operations (XOR and bit count) in network evaluation. It should be remarked that XOR can be evaluated efficiently in a garbled circuit protocol with negligible computation and zero communication. XONN reports 0.15 sec (resp. 5.79 sec) and 32.13 MB (resp.  2599 MB) at accuracy 99\% (resp. 81.85\%) for MNIST and CIFAR-10, respectively. 
}

\caesar{Table~\ref{tab:hal:sota} provides a high-level comparison between the state-of-the-art private inference frameworks. Pure MPC- and mixed MPC-FHE-based solutions suffer from non-constant number of interactions that depends on the number of non-linear layers in the neural network. XONN manages to avoid that as they adopt binarized networks which do not require multiplication and can be evaluated without interaction. On the other hand, pure FHE solutions require a single round of communication in which the client provides the input and FHE evaluation keys and receive the output after evaluation. It can also be noticed that, by design, MPC-based solutions require both the client and the server to know the network architecture due to the adoption of garbled circuits.   
}

\begin{table*}[htbp]
\footnotesize
  \centering
  \aboverulesep=0ex
 \belowrulesep=0ex
  \renewcommand{\arraystretch}{1.3}
  \caption{The hallmarks of privacy-preserving inference frameworks. NLL: non-linear layers; GC: garbled circuits; SS: secrete sharing.}
    \begin{tabular}{|l|c|c|c|c|c|c|c|c|}
    \toprule
    \multicolumn{1}{|c|}{\textbf{Framework}} & \multicolumn{1}{p{4.6em}|}{\textbf{$\#$ Rounds\newline{}Interaction}} & \multicolumn{1}{p{5.8em}|}{\textbf{Cryptographic\newline{}Tools}} & \multicolumn{1}{p{5.1em}|}{\textbf{Non-Linear Activation \newline{}Functions}} & \multicolumn{1}{p{5.5em}|}{\textbf{Disclosure of \newline{}Inference\newline{}Result}} & \multicolumn{1}{p{4.0em}|}{\textbf{Batched\newline{}Inference}} & \multicolumn{1}{p{5.7em}|}{\textbf{Offline \newline{}Preprocessing}} & \multicolumn{1}{p{7.5em}|}{\textbf{Training/Testing \newline{}Network Architecture}} & \multicolumn{1}{p{5.5em}|}{\textbf{Disclosure of \newline{}Network Architecture}} \\
    \midrule
    \midrule
    \textbf{MiniONN} & $1 + \# \text{ NLL}$  & \multicolumn{1}{p{6.355em}|}{GC, SS, \newline{}Additive HE} & Exact & Client, Server & No    & Low   & Same  & Client, Server \\
    \midrule
    \textbf{GAZELLE} & $1 + \# \text{ NLL}$ & \multicolumn{1}{p{6.355em}|}{GC, SS, \newline{}Additive HE} & Exact & Client, Server & No    & Medium & Same  & Client, Server \\
    \midrule
    \textbf{DELPHI} & $1 + \# \text{ NLL}$ & GC, SS & Polynomial & Client, Server & No    & High  & Same  & Client, Server \\
    \midrule
    \textbf{XONN} & 1 & GC, SS & Exact & Client, Server & No    & High  & Same  & Client, Server \\
    \midrule
    \textbf{CryptoNets} & 1     & HE    & Polynomial & Client only & Yes   & Low   & Different & Server only \\
    \midrule
    \textbf{FCryptoNets} & 1     & HE    & Polynomial & Client only & No   & Low   & Same & Server only \\
    \midrule
    \textbf{E2DM } & 1     & HE    & Polynomial & Client only & Yes   & Low   & Same  & Server only \\
    \midrule
    \textbf{DiNN} & 1     & HE    & Polynomial & Client only & No    & Low   & Same  & Server only \\
    \midrule
    \textbf{HCNN} & 1     & HE    & Polynomial & Client only & Yes   & Low   & Same  & Server only \\
    \bottomrule
    \end{tabular}%
  \label{tab:hal:sota}%
\end{table*}%

\subsection{Organization of the Paper}
Section~\ref{section:Preliminaries} introduces fully homomorphic encryption and NNs. Following that, Section~\ref{section:hcnn} discusses the challenges of adapting CNNs to the homomorphic domain.
Next, we describe the components that were used in implementing HCNNs in Section~\ref{section:implementation}.
In Section~\ref{section:experiments}, we report the results of experiments done using our implementation of HCNNs on MNIST and CIFAR-10 datasets.
Lastly, we conclude with Section~\ref{section:conclusion} and provide potential research directions.

\section{Preliminaries}\label{section:Preliminaries}

\subsection{Fully Homomorphic Encryption}\label{subsection:hePreliminaries}
First proposed by Rivest \etal~\cite{FOSC:RivAdlDer78}, (FHE) was envisioned to enable arbitrary computation on encrypted data.
FHE would support operations on ciphertexts that translate to functions on the encrypted messages within. 
It remained unrealized for more than 30 years until Gentry~\cite{STOC:Gentry09} proposed the first construction.
The blueprint of this construction remains the only method to design FHE schemes.
The (modernized) blueprint is a simple two-step process.
First, a somewhat homomorphic encryption scheme that can evaluate its decryption function is designed.
Then, we perform bootstrapping, which decrypts a ciphertext using an encrypted copy of the secret key.

As bootstrapping imposes high computation costs, we adopt a levelled FHE scheme instead, which can evaluate functions up to a pre-determined multiplicative depth without bootstrapping. We chose the Brakerski-Fan-Vercauteren~(BFV) scheme~\cite{C:Brakerski12,EPRINT:FanVer12}, whose security is based on the Ring Learning With Errors (RLWE) problem proposed by Lyubashevsky \etal~\cite{EC:LyuPeiReg10}.
This problem is conjectured to be hard even with quantum computers, backed by reductions (in~\cite{EC:LyuPeiReg10} among others) to worst-case problems in ideal lattices. 

The BFV scheme has five algorithms~(\textsf{KeyGen}, \textsf{Encrypt}, \textsf{Decrypt}, \textsf{HAdd}, \textsf{HMult}).
\textsf{KeyGen} generates the keys used in a FHE scheme given the parameters chosen.
\textsf{Encrypt} and \textsf{Decrypt} are the encyption and decryption algorithms respectively.
The differentiation between FHE and standard public-key encryption schemes is the operations on ciphertexts; which we call \textsf{HAdd} and \textsf{HMult}.
\textsf{HAdd} outputs a ciphertext that decrypts to the sum of the two input encrypted messages while \textsf{HMult} outputs one that decrypts to the product of the two encrypted inputs.

We informally describe the basic scheme below and refer to~\cite{EPRINT:FanVer12} for the complete details.
Let \(k, q, t > 1\) with \(N = 2^k\), \(t\) prime and \(\calR = \mathbb{Z}[X]/\langle X^{N} + 1\rangle\), we denote the ciphertext space as \(\calR_q = \calR/q\calR\) and message space as \(\calR_t = \calR/t\calR\).
We call ring elements ``small'' when their coefficients have small absolute value.
\begin{itemize}\itemsep=0.1mm
    \item \textsf{KeyGen}(\(\lambda, L\)): Given security parameter \(\lambda\) and level \(L\) as inputs, choose \(k, q\) so that security level \(\lambda\) is achieved.
    Choose a random element \(a \in \calR_q\), ``small'' noise \(e \in \calR_q\) and secret key \(s \in \calR_2\), the public key is defined to be \(pk = (b = e-as, a)\). We note that for relinearlization, an evaluation key (evk) is also generated to help control the size of ciphertext after homomorphic multiplications. First, choose an integer $w$ to control the decomposition rate and number of components $l+1$ in evk, where $l = \lfloor \log_w{q} \rfloor$. For $0 < i \leq l$, sample $a_i \in R_q$ and $e_i \in R_q$ and compute $evk[i] = ( [w^is^2 - (a_is+e_i)]_q, a_i)$. 
    \item \textsf{Encrypt}(\(pk, m\)): Given public key \(pk\) and message \(m \in \calR_t\) as input, the encryption of \(m\) is defined as \(\bfc = (br^{\prime} + e^{\prime} + \lfloor q/t\rfloor m, ar^{\prime})\), for some random noise \(e^{\prime}, r^{\prime} \in \calR_q\).
    \item \textsf{Decrypt}(\(sk, \bfc\)): Given secret key \(sk\) and ciphertext \(\bfc = (c_0, c_1) \in \calR^2_q\) as inputs, the decryption of \(\bfc\) is 
    \begin{equation*}
        m = \left\lceil (t/q) (c_0 + c_1s \bmod q) \right\rfloor \bmod t.
    \end{equation*}
    \item \textsf{HAdd}(\(\bfc_1, \bfc_2\)): Given two ciphertexts \(\bfc_1 = (c_{0,1}, c_{1,1}), \bfc_2 = (c_{0,2}, c_{1,2})\) as inputs, the operation is simply component-wise addition, i.e. the output ciphertext is \(\bfc^{\prime} = (c_{0,1} + c_{0,2}, c_{1,1} + c_{1,2})\).
    \item \textsf{HMult}(\(\bfc_1, \bfc_2\)): Given two ciphertexts \(\bfc_1 = (c_{0,1}, c_{1,1}), \bfc_2 = (c_{0,2}, c_{1,2})\) as inputs, proceed as follows:
    \begin{enumerate}
        \item (Tensor) compute 
        \begin{equation*}\label{equation:Tensor}
            \begin{aligned}
                \bfc^{\ast} = (c_0 &= c_{0,1}c_{0,2}, c_1 = c_{0,1}c_{1,2} + c_{1,1}c_{0,2}, \\ c_2 &= c_{1,1}c_{1,2});
            \end{aligned}
        \end{equation*}
        \item (Scale and Relinearize) output 
        \begin{equation*}\label{equation:Relinearize}
            \begin{aligned}
                \bfc^{\prime} = \left\lceil \mathsf{Relinearize}(\lceil (t/q)\bfc^{\ast} \rfloor) \right\rfloor \bmod q.
            \end{aligned}
        \end{equation*}
    \end{enumerate}
\end{itemize}
Where $\mathsf{Relinearize} (\bfc^{\ast})$ is used to shrink the size of $\bfc^{\ast}$ from three back to two terms and defined as follows: decompose $c_2$ in base $w$ as $c_2 = \sum_{i=0}^l c_2^{(i)} w^i$. Return $\mathbf{c}^{\prime} = c_j + \sum_{i=0}^l evk[i][j]c_2^{(i)}$, where $j \in \{0,1\}$.

\subsubsection{Computation Model with Fully Homomorphic Encryption} 
The set of functions that can be evaluated with FHE are arithmetic circuits over the plaintext ring \(\mathcal{R}_t\).
However, this is not an easy plaintext space to work with; elements in \(\mathcal{R}_t\) are polynomials of degree up to several thousand.
Addressing this issue, Smart and Vercauteren~\cite{DCC:SmaVer14} proposed a technique to support single instruction multiple data~(SIMD) by decomposing \(\mathcal{R}_t\) into a product of smaller spaces with the CRT over polynomial rings.
For prime \(t \equiv 1 \bmod 2N\), \(X^{N}+1 \equiv \prod_{i=1}^{N} (X - \alpha_i) \bmod t\) for some \(\alpha_i \in \{1,2,\ldots,t-1\}\).
This means that \(\mathcal{R}_t = \prod_{i=1}^{N}\mathbb{Z}_t[X]/\langle X-\alpha_i\rangle \cong \prod_{i=1}^{N} \mathbb{Z}_t\).
Therefore, the computation model generally used with homomorphic encryption is arithmetic circuits with modulo \(t\) gates.

For efficiency, the circuits evaluated using the \textsf{HAdd} and \textsf{HMult} algorithms should be levelled.
This means that the gates of the circuits can be organized into layers, with inputs in the first layer and output at the last, and the outputs of one layer are inputs to gates in the next layer.
In particular, the most important property of arithmetic circuits for FHE is its depth.
The depth of a circuit is the maximum number of multiplication gates along any path of the circuit from the input to output layers.

A levelled FHE scheme with input level \(L\) can evaluate circuits of at most depth \(L\), which affects the choice of parameter \(q\) due to noise in ciphertexts.
In particular, the \textsf{HMult} operation on ciphertext is the main limiting factor to homomorphic evaluations.
\subsection{Neural Networks}\label{subsection:nnPreliminaries}
A neural network can be seen as an arithmetic circuit comprising a certain number of layers. Each layer consists of a set of nodes, with the first being the input of the network.
Nodes in the layers beyond the first take the outputs from a subset of nodes in the previous layer and output the evaluation of an activation function over them.
The values of the nodes in the last layer are the outputs of the neural network.

The most widely-used layers are:
\begin{enumerate}\itemsep=0.1mm
    \item Activation layers: Each node in this layer takes the output, \(z\), of a single node of the previous layer and outputs \(f(z)\) \caesar{for some activation function \(f\).}
    \item Convolution-Type layers: Each node in this layer takes the outputs, \(\mathbf{z}\), of some subset of nodes from the previous layer and outputs a weighted-sum \(\langle \mathbf{w}, \mathbf{z} \rangle + b\) for some weight vector \(\mathbf{w}\) and bias \(b\).
    \item Pooling layers: Each node in this layer takes the outputs, \(\mathbf{z}\), of some subset of nodes from the previous layer and outputs \(f(\mathbf{z})\) for some function \(f\).
\end{enumerate}

The functions used in the activation layers are quite varied, including sigmoid~(\(f(z) = \frac{1}{1+e^{-z}}\)), softplus~(\(f(z) = \log(1+e^{z})\)) and \(\relu\), where $\relu(z) = z, \text{if } z \geq 0$ or $0$ if $z$ is negative.
To adapt NNs operations over encrypted data, we use the following layers:
\begin{itemize}\itemsep=0.1mm
    \item \textit{Convolution~(weighted-sum) layer}: at each node, we take a subset of the outputs of the previous layer, also called a filter, and perform a weighted-sum on them to get its output.
    
    \item \textit{Average-Pooling layer}: at each node, we take a subset of the outputs of the previous layer and compute the average on them to get its output.
    
    \item \textit{Square} layer: each node linked to a single node~\(z\) of the previous layer; its output is the square of \(z\)'s output.
    \item \textit{Fully Connected} layer: similar to the convolution layer, each node outputs a weighted-sum, but over the entire previous layer rather than a subset of it.
\end{itemize}

\section{Homomorphic CNNs}\label{section:hcnn}

Homomorphic encryption (HE) enables computation directly on encrypted data.
This is ideal to handle the challenges that DL face when it comes to questions of data privacy.
We call CNNs that operate over encrypted data as Homomorphic CNNs~(HCNNs).
Although FHE promises a lot, there are several challenges that prevent straightforward translation of standard techniques for traditional CNNs to HCNNs. These challenges are described below.

\subsection{Plaintext Space}\label{sec:plaintext:packing}
\begin{figure}[!ht]
    \centering
    \includegraphics[width=.28\textwidth, height=3cm]{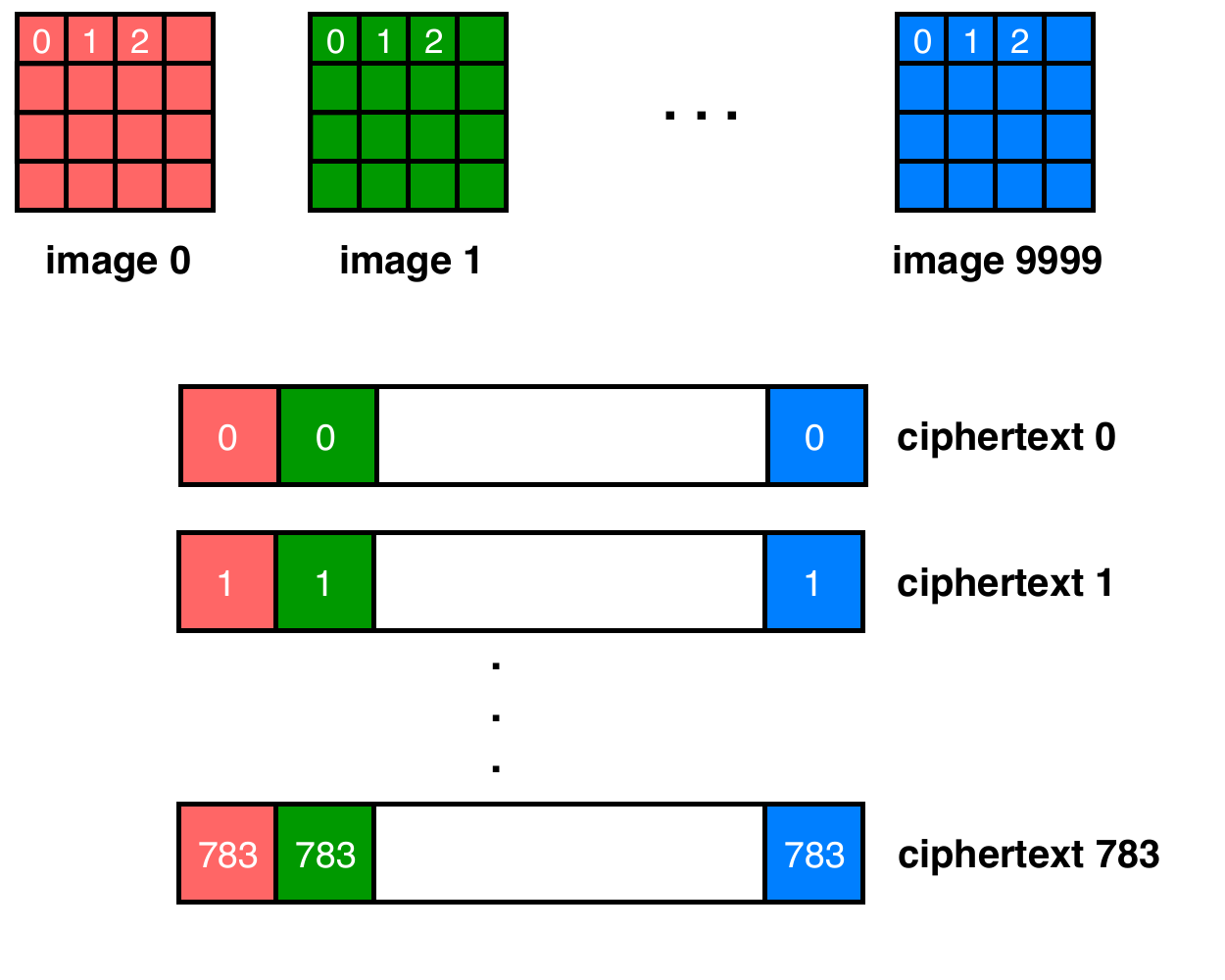}
    \caption{Packing MNIST testing dataset. Ciphertext $i$ contains pixel $i$ from all images.}
    \label{fig:mnist:packing}
\end{figure}

The first problem is the choice of plaintext space for HCNN computation.
Weights and inputs of a neural network are usually decimals, which are represented in floating-point.
Unfortunately, these cannot be directly encoded and processed in most FHE libraries and thus require special treatment. For simplicity and to allow inference on large datasets, we pack the same pixel of multiple images in a single ciphertext as shown in Figure~\ref{fig:mnist:packing}. This packing is useful for batched-inference scenarios (where the client has a large number of images to classify). For example, the client could be a hospital that needs to run a disease detector on a number of images that belong to different patients. Note that the BFV scheme can be instantiated such that a ciphertext may contain a certain number of slots to store multiple plaintext messages, i.e., ciphertexts can be viewed as vectors. We remark that this packing scheme was first proposed by CryptoNets~\cite{MSFT:DGL+16}.

\subsubsection{Encoding into the Plaintext Space} 
We adopt the \textit{scalar encoding}, which approximates these decimals with integers.
It is done by multiplying them with a scaling factor \(\Delta\) and rounding the result to the nearest integer.
Then, numbers encoded with the same scaling factor can be combined together using integer addition or multiplication.
For simplicity, we normalize the inputs and weights of HCNNs in between \([0,1]\) and \(\Delta\) (initially) corresponds to the number of bits of precision of the approximation, as well as the upper bound on the approximation.

Although straightforward to use, there are some downsides to this encoding.
The scale factor cannot be adjusted mid-computation and mixing numbers with different scaling factors is not straightforward.
For example, suppose we have two messages \(\Delta_1 m_1, \Delta_2 m_2\) with two different scaling factors, where \(\Delta_1 < \Delta_2\):
\begin{equation*}
    \begin{aligned}
        \Delta_1 m_1 + \Delta_2 m_2 &= \Delta_2 ( m_2 + \Delta_2/\Delta_1 m_1) \\ 
        \Delta_1 m_1 \times \Delta_2 m_2 &= \Delta_1\Delta_2 (m_1m_1).
    \end{aligned}
\end{equation*}
Multiplication will just change the scaling factor of the result to \(\Delta_1\Delta_2\) but the result of adding two encoded numbers is not their standard sum.
This means that as homomorphic operations are done on encoded data, the scaling factor in the outputs increases without a means to control it.
Therefore, the plaintext modulus \(t\) has to be large enough to accommodate the maximum number that is expected to result from homomorphic computations.

With the smallest scaling factor, \(\Delta = 2\), \caesar{a total of} \(64\) multiplications will suffice to cause the result to potentially overflow the space of \(64\)-bit integers.
Unfortunately, we use larger \(\Delta\) in most cases which means that the expected maximum will be much larger.
Thus, we require a way to handle large plaintext moduli of possibly several hundred bits, which is described next.

\begin{figure}[!ht]
    \centering
    \includegraphics[width=.40\textwidth, height=6.5cm]{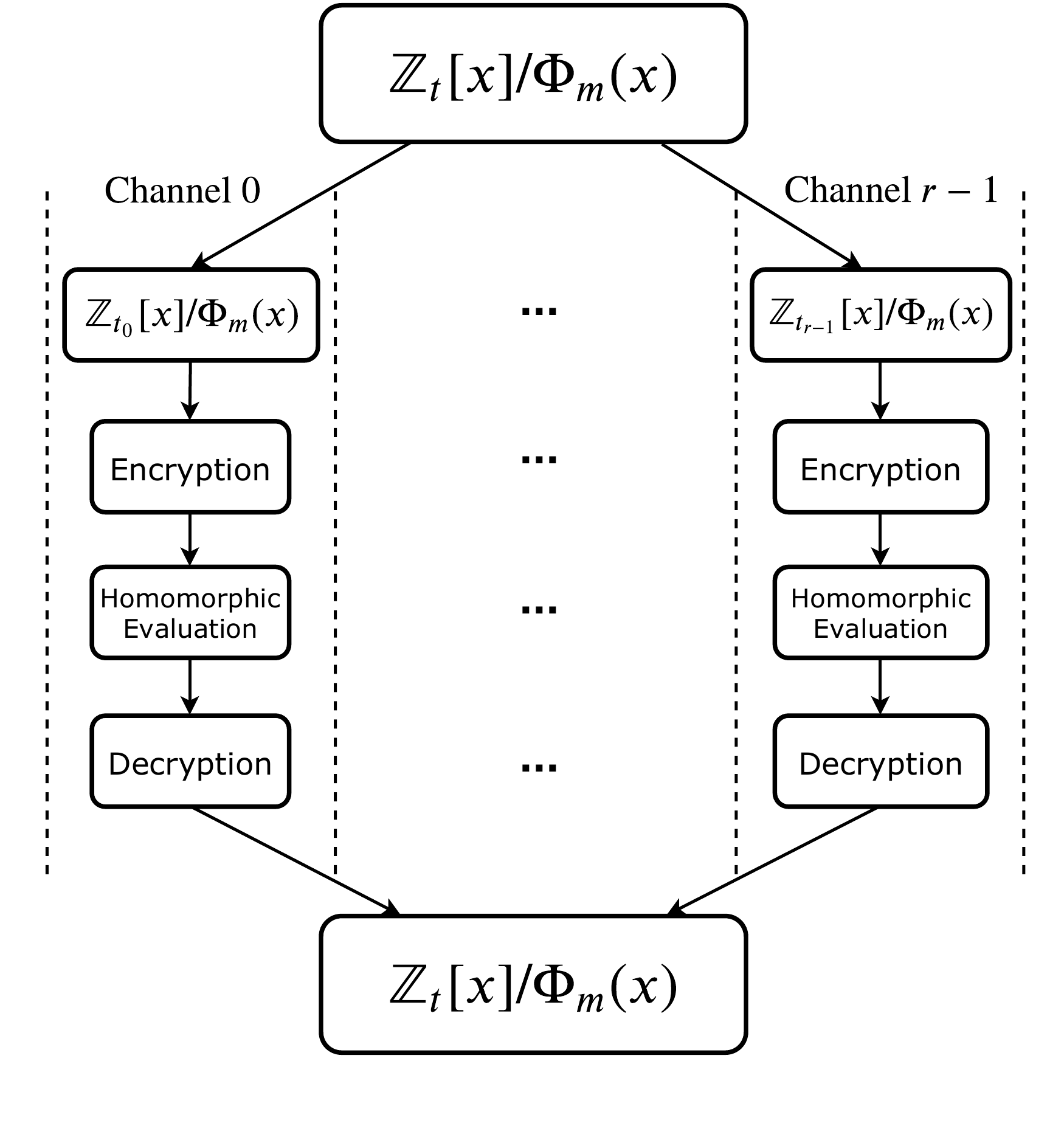}
    \caption{Plaintext Chinese Remainder Theorem (CRT) decomposition for a FHE arithmetic circuit}
    \label{fig:plaintext:crt:decomposition}
\end{figure}

\subsubsection{Plaintext Space CRT Decomposition}
One way to achieve this is to use a composite plaintext modulus, \(t = \prod_{i=0}^{r-1} t_i\) for some primes \(t_0, \ldots, t_{r-1}\) such that \(t\) is large enough to accommodate the maximum intermediate result the network may generate.
Recall that the CRT gives us an isomorphism between \(\mathbb{Z}_t\) and \(\prod_{i=0}^{r-1} \mathbb{Z}_{t_i}\):
\begin{equation*}
    \begin{aligned}
        \mathsf{CRT}:~\mathbb{Z}_{t_0} \times \cdots \times \mathbb{Z}_{t_{r-1}} &\longrightarrow \mathbb{Z}_t \\
        \mathbf{m} = (m_0,~\ldots,~m_{r-1}) &\longmapsto m,
        \end{aligned} 
 \end{equation*}
where $m_i \in \mathbb{Z}_{t_i}$ and $m \in \mathbb{Z}_{t}$. The inverse map is:

 \begin{equation*}
    \begin{aligned}
        \mathsf{ICRT}:~\mathbb{Z}_t &\longrightarrow \mathbb{Z}_{t_0} \times \cdots \times \mathbb{Z}_{t_{r-1}} \\
        m &\longmapsto \mathbf{m} = (m_0,~\ldots,~m_{r-1});
    \end{aligned}
\end{equation*}
where for any \(m \in \mathbb{Z}_t\), we have \(\mathsf{CRT}(\mathsf{ICRT}(m)) = m\).

For such moduli, we can decompose any integer \(m < t\) into a length-\(r\) vector with \(\mathsf{ICRT}\).
Arithmetic modulo \(t\) is replaced by component-wise addition and multiplication modulo the prime \(t_i\) for the \(i\)-th entry of $\mathbf{m}$.
We can recover the output of any computation with \(\mathsf{CRT}\).

As illustrated in Figure~\ref{fig:plaintext:crt:decomposition}, for homomorphic operations modulo \(t\), we separately encrypt each entry of \(\mathbf{m}\) in \(r\) FHE instances with the appropriate \(t_i\) and perform modulo \(t_i\) operations.
At the end of the homomorphic computation of function \(f\), we decrypt the \(r\) ciphertexts, which gives us \(f(\mathbf{m})\).
The actual output \(f(m)\) is obtained by applying the CRT map to \(f(\mathbf{m})\). 
The example below describes how CRT works. 
\begin{exmp}
Suppose FHE parameters are set to support $t = 3 \cdot 5 = 15$ to calculate $y = 2 x + 1$, where $x = 4$. We would like to use CRT and run two FHE instances with $t_0 = 3$ and $t_1 = 5$ to find the value of $y$, which is equals to $2 \cdot 4 + 1 = 9$. This can be done as follows:
$\textbf{x} = x \pmod{t_i} = (1, 4)$. We calculate $y$ for each instance as: $\textbf{y} = 2x_{t_i}+1 \pmod{t_i} = (0,4)$. To find $y$, we apply the CRT reconstruction map as follows:
$y = \sum_{i=0}^{1} (\textbf{y}_i * ((\frac{t}{t_i})^{-1}) \mod{t_i}) \cdot \frac{t}{t_i} = 9$.

\end{exmp}

\subsection{Neural Network Layers}\label{subsection:hcnnAdapt}
Computation in FHE schemes are generally limited to addition and multiplication operations over ciphertexts. 
As a result, it is easy to compute polynomial functions with FHE schemes. 
As with all FHE schemes, encryption injects a bit of noise into the data and each operation on ciphertexts increases the noise within it. 
As long as the noise does not exceed some threshold, decryption is possible. Otherwise, the decrypted results are essentially meaningless. 

\subsubsection{Approximating Non-Polynomial Activations}
For CNNs, a major stumbling block for translation to the homomorphic domain is the activation functions. 
These are usually not polynomials, and therefore unsuitable for evaluation with FHE schemes. 
The effectiveness of the \(\relu\) function in CNNs means that it is almost indispensable.
Therefore, it should be approximated by some polynomial function to try to retain as much accuracy as possible.
The choice of approximating polynomial depends on the desired performance of the HCNN.
For example, in this work, we applied the square function, \(z \mapsto z^2\), which Dowlin~\etal~\cite{MSFT:DGL+16} found to be sufficient for accurate results on the MNIST dataset with a five-layer network.

The choice of approximation polynomial affects the depth of the activation layers as well as its complexity~(number of \textsf{HMult}s).
The depth and complexity of this layer will be \(\lceil \log d\rceil\) and \(d-1\) respectively, where \(d\) is the degree of the polynomial\footnote{Note that using a binary tree structure, one can evaluate a degree $d$ polynomial, with a circuit of multiplicative depth $\log d$.}.
However, with the use of scalar encoding, there is another effect to consider.
Namely, the scaling factor on the output will be dependent on the depth of the approximation, i.e., if the scaling factor of the inputs to the activation layer is \(\Delta\), then the scaling factor of the outputs will be roughly \(\Delta^{1 + \lceil \log d \rceil}\), assuming that the approximation is a monic polynomial.

\subsubsection{Handling Pooling Layers} 
Similar to activations, the usual functions used in pooling layers, maximum~(\(\mathsf{max}(\mathbf{z}) = \max_{1 \leq i \leq n} z_i\)), \(\ell_2\)-norm and mean~(\(\mathsf{avg}(\mathbf{z}) = \frac{1}{n}\sum_{i=1}^{n} z_i\)) for inputs \(\mathbf{z} = (z_1, \ldots, z_n)\), are generally non-polynomial.
Among these, \(\mathsf{avg}\) is the most FHE-friendly as it requires a number of additions and scaling by a known constant. We note that several works~\cite{ICLR:SDBR15,JMIA:KLN+17} have shown that pooling is not strictly necessary and good results can be obtained without it. We found that pooling layers are not necessary for our MNIST network. On the other hand, pooling gave better accuracy results with the CIFAR-10 network.

\subsubsection{Convolution-Type Layers} 
Lastly, we have the convolutional-type layers.
Since these are weighted sums, they are straightforward to compute over encrypted data; the weights can be multiplied to encrypted inputs with \textsf{HMultPlain} and the results summed with \textsf{HAdd}.
Nevertheless, we still have to take care of the scaling factor of outputs from this layer.
At first thought, we may take the output scaling factor as \(\Delta_w\Delta_i\), multiply the scaling factor of the weights and the inputs, denoted with \(\Delta_w\) and \(\Delta_i\) respectively.
But, there is the potential for numbers to increase in bit-size from the additions done in weighted sums.
Recall that when adding two \(\Delta\)-bit numbers, the upper bound on the sum is \(\Delta+1\) bits long.
Therefore, the maximum number that can appear in the worst-case in the convolutions is about \(\Delta_w\Delta_i\times2^{\lceil\log n \rceil}\) bits long, where \(n\) is the number of terms in the summands.
In practice, this bound is usually not achieved since the summands are seldom all positive.
With negative numbers in the mix, the actual contribution from the summation can be moderated by some constant \(0 < c < 1\).

\section{Implementation}\label{section:implementation}

Implementation is comprised of two parts: 1) training on unencrypted data and 2) classifying encrypted data. Training is performed using the 5-layer (for MNIST) and 11-layer (for CIFAR-10) networks whose details are shown in Table~\ref{table:mnistInference} and~\ref{tab:cifar10Inference}, respectively. We use the Tensorpack framework~\cite{wu2016tensorpack} to train the network and compute the model. This part is quite straightforward and can be simply verified by classifying the unencrypted test dataset. For NNs design, one of the major constraints posed by FHE is the limitation of numerical precision of layer-wise weight variables. Training networks with lower precision weights would significantly prevent the precision explosion in ciphertext as network depth increases and thus speed up inference rate in encrypted domain. To this end, we propose to train low-precision networks from scratch, without incurring much loss in accuracy compared to networks trained in floating-point precision. Following~\cite{zhou2016dorefa}, for each convolutional layer, we quantize floating point weight variables $w$ to $k$ bits numbers $w_q$ using simple uniform scalar quantizer shown below:
\begin{equation*}
        w_q = \frac{1}{2^k-1}round(w * (2^k-1))
\label{eq:quant}
\end{equation*}
This equation is non-differentiable, we use Straight Through Estimator (STE)~\cite{BengioLC13} to enable the back-propagation\footnote{STE is a technique that can be used to train quantized NNs which require differentiating piece-wise constant functions. It simply replaces these non-differentiable functions with identity functions in back-propagation.}. We trained the 5-layer network on MNIST training set with a precision of weights at 2, 4, 8 and 32 bits, and evaluated on MNIST test set with reported accuracy 96\%, 99\%, 99\% and 99\% respectively. In view of this, we choose the 4-bit network for the following experiments. It's worth noting that CryptoNets~\cite{MSFT:DGL+16} requires 5 to 10 bits of precision on weights to hit 99\% accuracy on MNIST test set, while our approach further reduces it to 4 bits and still maintains the same accuracy with both floating-point and scalar encoding. For CIFAR-10, as the problem is more challenging and the network is deeper, our network shown in Table~\ref{tab:cifar10Inference} uses 8 bits and achieves 77.80\% and 77.55\% classification accuracy with the floating-point and scalar encoding, respectively. We remark that the highest state-of-the-art accuracy of MNIST and CIFAR-10 in the unencrypted domain \caesar{(i.e., plaintext samples)} is 99.79\%~\cite{wan2013regularization} and 96.53\%~\cite{graham2014fractional}, respectively.

The second part is more involved as it requires running the network (with the pre-learned model) on encrypted data. First, we need to fix FHE parameters to accommodate for both the network multiplicative depth and precision. We optimized the scaling factors in all aspects of the HCNN. For the MNIST network, inputs were normalized to \([0,1]\), scaled by \(4\) and then rounded to the nearest integer. With the low-precision network trained from scratch, we convert the weights of the convolution-type layers to short \(4\)-bit integers, using a small scaling factor of \(15\); no bias was used in the convolutions. Similarly, inputs to the CIFAR-10 network were normalized to \([0,1]\) but we used much larger scale factors for the convolution layers. Moreover, padding has been used as it provided better accuracy.
The scaling factors for both networks are shown in Tables~\ref{table:mnistInference} and~\ref{tab:cifar10Inference}. 

Next, we implement the networks (with scalar encoding) using NTL~\cite{shoup2005ntl} (a multi-precision number theory C++ library). NTL is used to facilitate the treatment of the scaled inputs and accommodate for precision expansion of the intermediate values during the network evaluation. We found that the largest precision needed is less than ($2^{43}$) for MNIST and ($2^{218}$) for CIFAR-10. Note that for MNIST, it is low enough to fit in a single word on 64-bit platforms without overflow. On the other hand, we use the plaintext CRT decomposition to handle the large plaintext modulus required for CIFAR-10. By estimating the maximum precision required by the networks, we can estimate the FHE parameters required by HCNN.
\begin{table}[t]
    \scriptsize
    \centering
    \renewcommand{\arraystretch}{1.8}
    \caption{HCNN architecture for training and testing MNIST dataset with the scale factor used in scalar encoding. Inputs are scaled by 4.}
    \begin{tabular}{|c|p{3cm}|c|c|}
        \hline
        \textsc{Layer Type} & \multicolumn{1}{c|}{\textsc{Description}} & \textsc{Layer Size} & \textsc{Scale}\\ \hline\hline
        \multirow{2}{*}{Convolution} & \multirow{2}{3.1cm}{\(5\) filters of size \(5 \times 5\) and stride \((2,2)\) without padding.} & \multirow{2}{*}{\(12 \times 12 \times 5\)} & \multirow{2}{*}{15} \\ 
        & & & \\ \hline
        
        \multirow{2}{*}{Square} & \multirow{2}{3cm}{Outputs of the previous layer are squared.} & \multirow{2}{*}{\(12 \times 12 \times 5\)} & \multirow{2}{*}{1} \\ & & & \\ \hline
        
        \multirow{2}{*}{Convolution} & \multirow{2}{3.1cm}{\(50\) filters of size \(5 \times 5\) and stride \((2,2)\) without padding.} & \multirow{2}{*}{\(4 \times 4 \times 50\)} & \multirow{2}{*}{15}\\ 
        & & & \\ \hline
        
        \multirow{2}{*}{Square} & \multirow{2}{3cm}{Outputs of the previous layer are squared.} & \multirow{2}{*}{\(4 \times 4 \times 50\)} & \multirow{2}{*}{1} \\ & & & \\ \hline
        
        \multirow{4}{*}{Fully Connected} & \multirow{4}{3.1cm}{Weighted sum of the entire previous layer with \(10\) filters, each output corresponding to \(1\) of the possible \(10\) digits.} & \multirow{4}{*}{\(1 \times 1 \times 10\)} & \multirow{4}{*}{15} \\ 
         & & & \\
         & & & \\
         & & & \\ \hline
    \end{tabular}
    \label{table:mnistInference}
\end{table}
\begin{table}[htbp]
    \scriptsize
    \centering
    \renewcommand{\arraystretch}{1.5}
    \caption{HCNN architecture for training and testing CIFAR-10 dataset with the scale factor used in scalar encoding. Inputs are scaled by 255.}
    \begin{tabular}{|c|p{3cm}|c|c|}
        \hline
        \textsc{Layer Type} & \multicolumn{1}{c|}{\textsc{Description}} & \textsc{Layer Size} & Scale \\ \hline \hline
        \multirow{2}{*}{Convolution} & \multirow{2}{3.1cm}{32 filters of size 3 x 3 x 3 and stride (1, 1) with padding.} & \multirow{2}{*}{32 x 32 x 32} & \multirow{2}{*}{10000}\\ 
        &       &  &  \\ \hline
        \multirow{2}{*}{Square} & \multirow{2}{3.1cm}{Outputs of the previous layer are squared.} & \multirow{2}{*}{32 x 32 x 32} & \multirow{2}{*}{1}\\ 
        &       &  &  \\ \hline
        \multirow{2}{*}{Pooling} & \multirow{2}{3cm}{Average pooling with extent 2 and stride 2.} & \multirow{2}{*}{16 x 16 x 32} & \multirow{2}{*}{4}\\ 
        &       &  &  \\ \hline
        \multirow{2}{*}{Convolution} & \multirow{2}{3.1cm}{64 filters of size 3 x 3 x 32 and stride (1, 1) with padding.} & \multirow{2}{*}{16 x 16 x 64} & \multirow{2}{*}{4095} \\ 
        &       &  &  \\ \hline
        \multirow{2}{*}{Square} & \multirow{2}{3cm}{Outputs of the previous layer are squared.} & \multirow{2}{*}{16 x 16 x 64} & \multirow{2}{*}{1}\\ 
        &       &  &  \\ \hline
        \multirow{2}{*}{Pooling} & \multirow{2}{3cm}{Average pooling with extent 2 and stride 2.} & \multirow{2}{*}{8 x 8 x 64} & \multirow{2}{*}{4}\\ 
        &       &  &  \\ \hline
        \multirow{2}{*}{Convolution} & \multirow{2}{3.1cm}{128 filters of size 3 x 3 x 64 and stride (1, 1) with padding.} & \multirow{2}{*}{8 x 8 x 128} & \multirow{2}{*}{10000}\\ 
        &       &  &  \\ \hline
        \multirow{2}{*}{Square} & \multirow{2}{3cm}{Outputs of the previous layer are squared.} & \multirow{2}{*}{8 x 8 x 128} & \multirow{2}{*}{1}\\ 
        &       &  &  \\ \hline
        \multirow{2}{*}{Pooling} & \multirow{2}{3cm}{Average pooling with extent 2 and stride 2.} & \multirow{2}{*}{4 x 4 x 128} & \multirow{2}{*}{4}\\ 
        &       &  &  \\ \hline
        \multirow{2}{*}{Fully Connected} & \multirow{2}{3.1cm}{Weighted sum of the entire previous layer with 256 filters} & \multirow{2}{*}{1 x 1 x 256} & \multirow{2}{*}{1023}\\
        &       &  &  \\  \hline
        \multirow{2}{*}{Fully Connected} & \multirow{2}{3cm}{Weighted sum of the entire previous layer with 10 filters.} & \multirow{2}{*}{1 x 1 x 10} & \multirow{2}{*}{63}\\
        &       &  &  \\ \hline
    \end{tabular}%
    \label{tab:cifar10Inference}%
\end{table}%

The next step is to implement the networks using a FHE library. We implement MNIST HCNN using two FHE libraries: SEAL~\cite{SEAL} and GPU-accelerated BFV (\afv) that is described in~\cite{TCHES:BVMA18}. On the other hand, we implement CIFAR-10 HCNN only using \afv~as it is more computationally intensive and would take a very long time on CPU. The purpose of implementing MNIST HCNN in SEAL is to facilitate a more unified comparison under the same system parameters and show the superiority of the GPU implementation. Also, we would like to highlight a limitation in the Residue Number Systems (RNS) variant that is currently implemented in SEAL. 

\subsection{HCNN Complexity}
In this section, we break down the complexity of both HCNNs layers and calculate the total number of operations required for homomorphic evaluation. We also compare our MNIST HCNN with \caesar{CryptoNets'} network~\cite{MSFT:DGL+16}.

Table~\ref{table:hcnn:complexity} shows the computational complexity of each layer in MNIST HCNN and CryptoNets. The convolution and fully connected layers require homomorphic multiplication of ciphertext by plaintext (\textsf{HMultPlain}). Suppose the input to the convolution or fully connected layers is vector $\textbf{i}$ of length $l$ and the output is vector $\textbf{o}$ of length $m$. Let $f_w$ and $f_h$ denote the filter width and height, respectively. The total number of \textsf{HMultPlain} in the convolution layer can be found by $m \cdot f_w \cdot f_h$. The fully connected layer requires $l \cdot m$ \textsf{HMultPlain} operations. On the other hand, the square layer requires $l = m$ ciphertext by ciphertext multiplications (\textsf{HMult}). It should be noted that most of FHE libraries provide an additional procedure for homomorphic squaring (\textsf{HSquare}) which has slightly lower computational complexity compared to \textsf{HMult}, (see Table~\ref{table:benchmarks}). It can be noticed that HCNN requires much lower number of \textsf{HMultPlain} compared to CryptoNets (46,000 vs 106,625). In CryptoNets, the third layer combines 4 linear layers (2 scaled mean pool, convolution and fully connected layers) for efficiency reasons, whereas it is simply a convolution layer in HCNN. On the other hand, CryptoNets requires less \textsf{HSquare} (945 vs 1,520). Note that MNIST HCNN requires a smaller plaintext modulus (43-bit) compared to CryptoNets (80-bit). This allows running a single instance of MNIST HCNN without plaintext decomposition, i.e., single CRT channel, whereas CryptoNets use higher precision training and require plaintext modulus of higher precision ($2^{80}$). This gives our HCNN at least 2$\times$ speedup against CryptoNets under unified system settings.

Table~\ref{table:hcnn:complexity:cifar10} shows the computational complexity for CIFAR-10 HCNN. It can be clearly seen that CIFAR-10 HCNN is more computationally intensive compared to MNIST. For instance, 6,952,332 \textsf{HMultPlain} and 57,344 \textsf{HMult} operations are required compared to 46,000 and 1,520, respectively for MNIST HCNN.
\begin{table*}[htbp]
    \scriptsize
  \centering
  \caption{MNIST HCNN vs CryptoNets~\cite{MSFT:DGL+16} complexity for homomorphic inference}
    \resizebox{\linewidth}{!}{\begin{tabular}{ccccccccc}
    \toprule
          & \multicolumn{4}{c}{MNIST HCNN} & \multicolumn{4}{c}{CryptoNets} \\
\cmidrule(lr){2-5} \cmidrule(lr){6-9}    
\multicolumn{1}{c}{\multirow{2}[4]{*}{Layer}} & Input Neurons & \multicolumn{1}{c}{Output Neurons} & \multicolumn{2}{c}{\# of Multiplications} & Input Neurons & \multicolumn{1}{c}{Output Neurons} & \multicolumn{2}{c}{\# of Multiplications} \\
\cmidrule{4-5} \cmidrule{8-9}          & \multicolumn{1}{r}{} &       & \multicolumn{1}{c}{\textsf{HMultPlain}} & \multicolumn{1}{c}{\textsf{HMult}} & \multicolumn{1}{r}{} &       & \multicolumn{1}{c}{\textsf{HMultPlain}} & \multicolumn{1}{c}{\textsf{HMult}} \\
    \midrule
    1     & 28$\times$28 = 784 & \multicolumn{1}{c}{5$\times$12$\times$12 = 720} & \multicolumn{1}{c}{25$\times$720 = 18,000} & \multicolumn{1}{c}{-} & 29$\times$29 = 841 & \multicolumn{1}{c}{5$\times$13$\times$13 = 845} & \multicolumn{1}{c}{25$\times$845 = 21,125} & \multicolumn{1}{c}{-} \\
    2     & 5$\times$12$\times$12 = 720 & \multicolumn{1}{c}{5$\times$12$\times$12 = 720} & \multicolumn{1}{c}{-} & 720   & 5$\times$13$\times$13 = 845 & \multicolumn{1}{c}{5$\times$13$\times$13 = 845} & \multicolumn{1}{c}{-} & 845 \\
    3     & 5$\times$12$\times$12 = 720 & \multicolumn{1}{c}{4$\times$4$\times$50 = 800} & \multicolumn{1}{c}{25$\times$800 = 20,000} & \multicolumn{1}{c}{-} & 5$\times$13$\times$13 = 845 & 1$\times$1$\times$100 = 100   & \multicolumn{1}{c}{100$\times$845 = 84,500} & \multicolumn{1}{c}{-} \\
    4     & 4$\times$4$\times$50 = 800 & \multicolumn{1}{c}{4$\times$4$\times$50 = 800} & \multicolumn{1}{c}{-} & 800   & \multicolumn{1}{c}{1$\times$1$\times$100 = 100} & 1$\times$1$\times$100 = 100   & \multicolumn{1}{c}{-} & 100 \\
    5     & 4$\times$4$\times$50 = 800 & \multicolumn{1}{c}{1$\times$1$\times$10 = 10} & \multicolumn{1}{c}{10$\times$800 = 8,000} & \multicolumn{1}{c}{-} & \multicolumn{1}{c}{1$\times$1$\times$100 = 100} & 1$\times$1$\times$10 = 10    & \multicolumn{1}{c}{10$\times$100 = 1,000} & \multicolumn{1}{c}{-} \\
    \midrule
          & \multicolumn{2}{p{16.425em}}{Total} & \textbf{46,000} & 1,520  & \multicolumn{2}{p{16.425em}}{Total} & 106,625 & \textbf{945} \\
    \bottomrule
    \end{tabular}}%
  \label{table:hcnn:complexity}%
\end{table*}%
\begin{table}[htbp]
    \scriptsize
    \centering
    \caption{CIFAR-10 HCNN complexity for homomorphic inference}
    \resizebox{\linewidth}{!}{\begin{tabular}{cp{9.355em}ccc}
        \toprule
        & \multicolumn{4}{c}{CIFAR-10 HCNN} \\
        \cmidrule{2-5}    \multicolumn{1}{c}{\multirow{2}[4]{*}{Layer}} & Input Neurons & \multicolumn{1}{c}{Output Neurons} & \multicolumn{2}{c}{No. of Multiplications} \\
        \cmidrule{4-5}          & \multicolumn{1}{r}{} &       & \multicolumn{1}{c}{\textsf{HMultPlain}} & \multicolumn{1}{c}{\textsf{HMult}} \\
        \midrule
        1     & 32 x 32 x 3 = 3072 & \multicolumn{1}{p{10.215em}}{32 x 32 x 32 = 32768} & 589,824 & \multicolumn{1}{c}{-} \\
        2     & 32 x 32 x 32 = 32768 & \multicolumn{1}{p{10.215em}}{32 x 32 x 32 = 32768} & \multicolumn{1}{c}{-} & 32,768 \\
        3     & 32 x 32 x 32 = 32768 & \multicolumn{1}{p{10.215em}}{16 x 16 x 32 = 8192} & 0     & \multicolumn{1}{c}{-} \\
        4     & 16 x 16 x 32 = 8192 & \multicolumn{1}{p{10.215em}}{16 x 16 x 64 = 16384} & 2,594,048 & \multicolumn{1}{c}{-} \\
        5     & 16 x 16 x 64 = 16384 & \multicolumn{1}{p{10.215em}}{16 x 16 x 64 = 16384} & \multicolumn{1}{c}{-} & 16,384 \\
        6     & 16 x 16 x 64 = 16384 & \multicolumn{1}{p{10.215em}}{8 x 8 x 64 = 4096} & 0     & \multicolumn{1}{c}{-} \\
        7     & 8 x 8 x 64 = 4096 & \multicolumn{1}{p{10.215em}}{8 x 8 x 128 = 8192} & 3,308,544 & \multicolumn{1}{c}{-} \\
        8     & 8 x 8 x 128 = 8192 & \multicolumn{1}{p{10.215em}}{8 x 8 x 128 = 8192} & \multicolumn{1}{c}{-} & 8,192 \\
        9     & 8 x 8 x 128 = 8192 & \multicolumn{1}{p{10.215em}}{4 x 4 x 128 = 2048} & 0     & \multicolumn{1}{c}{-} \\
        10    & 4 x 4 x 128 = 2048 & \multicolumn{1}{p{10.215em}}{1 x 1 x 256 = 256} & 457,398 & \multicolumn{1}{c}{-} \\
        11    & 1 x 1 x 256 = 256 & \multicolumn{1}{p{10.215em}}{1 x 1 x 10 = 10} & 2518  & \multicolumn{1}{c}{-} \\
        \midrule
        & \multicolumn{2}{p{18.57em}}{Total} & 6,952,332 & 57,344 \\
        \bottomrule
    \end{tabular}}%
    \label{table:hcnn:complexity:cifar10}%
\end{table}%

\subsection{Choice of Parameters}\label{subsec:parameters:choice}
Similar to other cryptographic schemes, one needs to select FHE parameters to ensure that known attacks are computationally infeasible. We denote to the desired security parameter by $\lambda$ measured in bits. This means that an adversary needs to perform $2^{\lambda}$ elementary (or bit) operations to break the scheme with probability one. A widely accepted estimate for $\lambda$ in the literature is $\geq$ 80 bits~\cite{smart2014algorithms}, which is used here to generate the BFV parameters.

In this work, we used a levelled BFV scheme that can be configured to support a known multiplicative depth $L$, which can be controlled by three parameters: $q$, $t$ and noise growth. The first two are problem dependent whereas noise growth depends on the scheme. As mentioned in the previous section, we found that $t$ should be at least 43 (resp. 218) bit integer for MNIST (resp. CIFAR-10) to accommodate the precision expansion in HCNN evaluation.

For our HCNNs, 5 (resp. 11) multiplicative depth is required: 2 (resp. 3) ciphertext by ciphertext (in the square layers) and 3 (resp. 8) ciphertext by plaintext (in convolution, pooling and fully connected layers) operations for MNIST and CIFAR-10 HCNNs, respectively. It is known that the latter has a lower effect on noise growth. This means that $L$ needs not to be set to 5 for MNIST and 11 for CIFAR-10. Experimentally, we found that $L = $ 4 and 7 are sufficient to run successfully MNIST and CIFAR-10 HCNNs, respectively in \afv. However, SEAL required higher depth ($L=5$) to run our MNIST HCNN. The reason is that SEAL implements the Bajard-Enyard-Hasan-Zucca (BEHZ)~\cite{bajard2016full} RNS variant of the BFV scheme that slightly increases the noise growth due to approximated RNS operations. Whereas in \afv, the Halevi-Polyakov-Shoup (HPS)~\cite{EPRINT:HalPolSho18} RNS variant is implemented, which has a lower effect on the noise growth~\cite{EPRINT:BPAVR18}.\\
Having $L$ and $t$ fixed, we can estimate $q$ using the noise growth bounds enclosed with the BFV scheme. 
Next, we try to estimate $N$ to ensure a certain security level. To calculate the security level, we used the LWE hardness estimator in~\cite{albrecht2015concrete} (commit \textsf{76d05ee}).\\
The above discussion suggests that the design space of HCNN is not limited depending on the choice of the plaintext coefficient modulus $t$. We identify a set of possible designs that fit different requirements. The designs vary in the number of factors in $t$ (i.e., number of CRT channels) and the provided security level. We provide four sets of parameters for MNIST and one for CIFAR-10. Parameter sets 2 and 4 are shown here to enable running our MNIST HCNN with SEAL. As will be shown later in the subsequent section, SEAL requires a higher $q$ due to the higher noise growth in its underlying FHE scheme.
Table~\ref{table:parameters} shows the system parameters used for each HCNN with the associated security level. Note that we use 10 primes for the plaintext moduli, each of size 22/23 bits, to run the CIFAR-10 HCNN. The product of these small primes gives us a 219-bit plaintext modulus which is sufficient to accommodate any intermediate result in CIFAR-10 HCNN. Note that we need to run 10 instances of CIFAR-10 HCNN to obtain the desired result. 
\caesar{ However, these instances are independent and can be run simultaneously (see Figure~\ref{fig:plaintext:crt:decomposition}).
}

It is worth noting also that the choice of the primes in the plaintext modulus is not arbitrary especially if one wants to do the inference for multiple images at once. To enable the packed encoding described in Section~\ref{sec:plaintext:packing}, one has to ensure that $2N|(t-1)$.
%
%

\begin{table}[t]
    \scriptsize
    \renewcommand{\arraystretch}{1.8}
    \centering
    \caption{HE parameters for MNIST and CIFAR-10 HCNNs with different parameter sets. Depth refers to the supported multiplicative depth and $\lambda$ denotes the security level in bits.}
    \begin{tabular}{lclcp{2.5cm}cr}
        \hline
        \multicolumn{1}{c}{HCNN}&
        \multicolumn{1}{c}{ID}&
        \multicolumn{1}{c}{$N$} & \multicolumn{1}{c}{$\log q$} & \multicolumn{1}{c}{Plaintext moduli} & 
        \multicolumn{1}{c}{Depth} & \multicolumn{1}{c}{$\lambda$} \\
        \hline
        \multirow{3}{*}{MNIST} & 1 & $2^{13}$ & 330 & 5522259017729 & 4 & 82 \\
        & 2 & $2^{13}$ & 360 & 5522259017729 & 5 & 76 \\
        & 3 & $2^{14}$ & 330 & 5522259017729 & 4 & 175 \\
        & 4 & $2^{14}$ & 360 & 5522259017729 & 5 & 159 \\
        \multirow{3}{*}{CIFAR-10} & \multirow{3}{*}{5} & \multirow{3}{*}{$2^{13}$} & \multirow{3}{*}{300} & \multirow{3}{3.1cm}{2424833, 2654209, 2752513, 3604481, 3735553, 4423681, 4620289, 4816897, 4882433, 5308417} & \multirow{3}{*}{7} & \multirow{3}{*}{91}\\
        & & & & & \\
        & & & & & \\
        \hline
    \end{tabular}
    \label{table:parameters}
\end{table}
\subsection{HCNN Inference Library}
As most DL frameworks do not use functions that fit the restrictions of FHE schemes, we designed an inference library using standard C++ libraries that implement some of the CNN layers using only additions and multiplications.
Support for arbitrary scaling factors per layer is included for flexibility and allows us to easily define neural network layers for HCNN inference.
We give a summary of the scaling factor growth of the layers we used in Table~\ref{table:scalingFactorGrowth}.
\begin{table}[!ht]
    \scriptsize
    \centering
    \renewcommand{\arraystretch}{1.5}
    \caption{Scaling Factor Growth by Layer}
    \begin{tabular}{ll}
        \toprule
        Layer Type & Output Scaling Factor \\ 
        \midrule
        Convolution-Type~(\(\sum_{i=1}^{n} w_iz_i\)) &  \(\Delta_o = \Delta_w\Delta_i\cdot2^{c\lceil \log n \rceil}\), \(0 < c < 1\). \\ 
        Square Activation~(\(f(z) = z^2\)) & \(\Delta_o = \Delta_i^{2}\). \\ \hline
        \multicolumn{2}{l}{where \(\Delta_i\) and \(\Delta_w\) are the input and weight scaling factors respectively.}
    \end{tabular}
    \label{table:scalingFactorGrowth}
\end{table}
In Section~\ref{subsection:nnPreliminaries}, we introduced several types of layers that are commonly used in designing NNs, namely activation, convolution-type and pooling.
Now, we briefly describe how our library realizes these layers.
For convolution-type layers, they are typically expressed with matrix operations but only require scalar additions and multiplications.
Our inference library implements them using the basic form, \(b + \sum^{n}_{i=1} w_i\cdot z_i\), for input \(\mathbf{z} = (z_1, \ldots, z_n)\) and weights \(\mathbf{w} = (w_1, \ldots, w_n)\). 

For the activation, some modifications had to be done for compatibility with FHE schemes.
In activation layers, the most commonly used functions are \(\relu\), sigmoid~(\(f(z) = \frac{1}{1+e^{-z}}\)) and softplus~(\(f(z) = \log(1+e^{z})\)).
These are non-polynomial functions and thus cannot be directly evaluated over FHE encrypted data.
Our library uses integral polynomials to approximate these functions; particularly for our HCNN, we used the square function, \(f(z) = z^2\), as a low-complexity approximation of \(\relu\).

The pooling layers used are average-pooling and they are quite similar to the convolution layers except that the weights are all ones and a scale factor that is equal to the reciprocal of the number of averaged values.

\subsection{GPU-Accelerated Homomorphic Encryption}
The FHE engine includes an implementation of an RNS variant of the BFV scheme~\cite{EPRINT:HalPolSho18} that is described in~\cite{TCHES:BVMA18, EPRINT:BPAVR18}. 
The BFV scheme is considered among the most promising FHE schemes due to its simple structure and low overhead primitives compared to other schemes. 
Moreover, it is a scale-invariant scheme where the ciphertext coefficient modulus is fixed throughout the entire computation. 
This contrasts to other scale-variant schemes that keep a chain of moduli and switch between them during computation. 
We use a GPU-based BFV implementation as an underlying FHE engine to perform the core FHE primitives: key generation, encryption, decryption and homomorphic operations such as addition and multiplication.

\begin{figure}[!h]
    \centering
    \includegraphics[width=.35\textwidth, height=4cm]{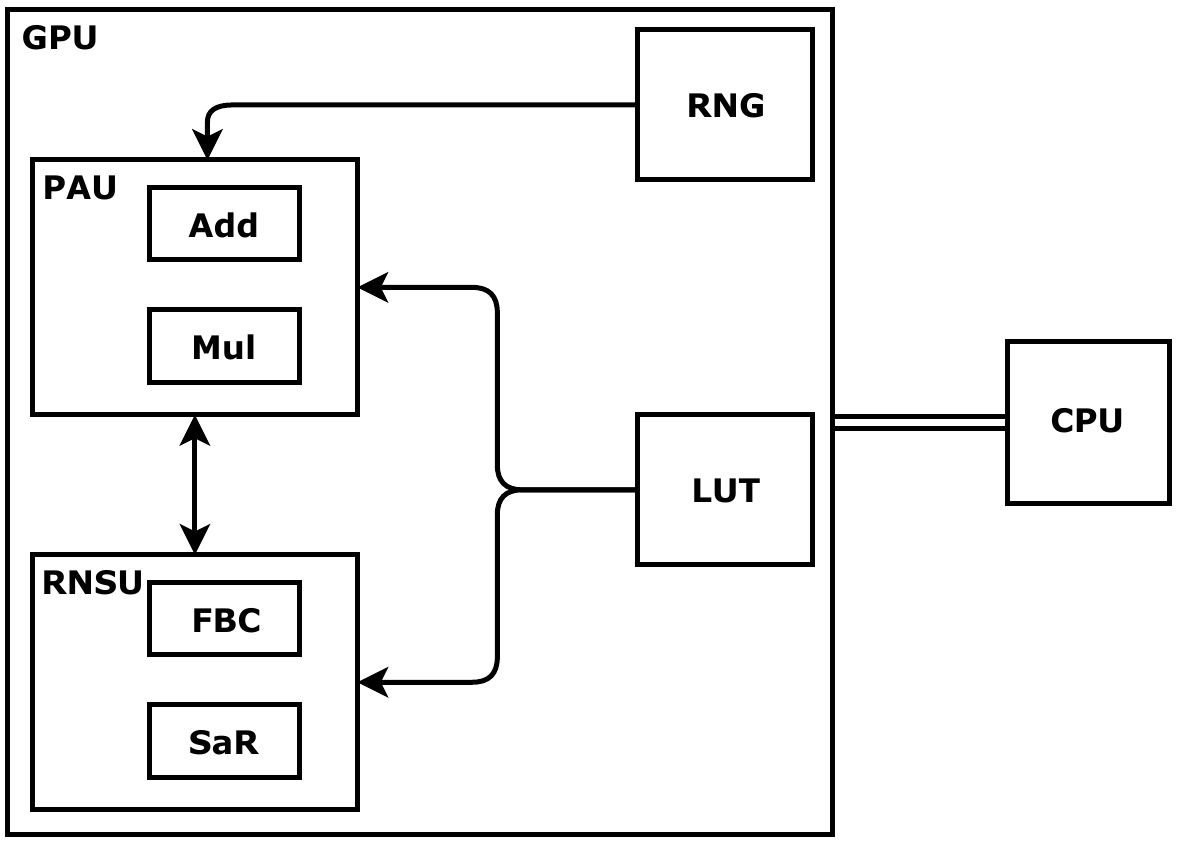}
    \caption{Top-Level Structure of the GPU-Accelerated \afv~Crypto-Processor.}
    \label{fig:bfv:gpu:processor}
\end{figure}

Our FHE engine (shown in Figure~\ref{fig:bfv:gpu:processor}) is comprised of three main components:  
\begin{enumerate}
    \item Polynomial Arithmetic Unit (PAU): performs basic polynomial arithmetic such as addition and multiplication.
    \item Residue Number System Unit (RNSU): provides additional RNS tools for efficient polynomial scaling required by BFV homomorphic multiplication and decryption.
    \item Random Number Generator Unit (RNG): used to generate random polynomials required by BFV key generation and encryption. 
\end{enumerate}
In addition, the FHE engine includes a set of Look-Up Tables (LUTs) that are used for fast modular arithmetic and number theoretic transforms required by both PAU and RNSU. For further details on the GPU implementation of BFV, we refer the reader to the referenced works.

\subsubsection{Fitting Large HCNN in GPU Memory}
Another major challenge facing FHE applications is their large memory requirements due to data expansion after encryption. The ciphertext size can be estimated as $2*N*\log_2{}q$ bits, this is approximately 1.28 MB and 0.59 MB for MNIST and CIFAR-10 HCNNs parameters, respectively. Moreover, due to homomorphic multiplication, ciphertext size may expand 3$\times$ its original size due to intermediate computation~\cite{EPRINT:BPAVR18}. As we encrypt one pixel - from multiple images - in one ciphertext, this means that in our HCNNs the number of ciphertexts for each layer is equal to the number of pixels in the feature map. For MNIST HCNN, the GPU memory (16 GB) was sufficient to store or generate all the ciphertexts in each layer. On the other hand, GPU memory was not enough to store the feature map of even the first convolution layer in CIFAR-10 HCNN. Therefore, we had to decompose the computation task into smaller subtasks, transfer them to GPU memory for execution, and finally transferring the result back to CPU memory. To do this, we have to decompose the 3 main computations: 1) convolution, 2) pooling, and 3) square. The square computation is straightforward to decompose as the input vector can be partitioned into shorter vectors and processed on GPU linearly. More challenging are convolution and pooling tasks which require a more sophisticated decomposition algorithm. To this end, we devise a decomposition method to fit the computation of both convolution and pooling in GPU memory. As pooling is a special case of convolution (the filter weights are all ones), we limit our discussion below to convolution.

\begin{figure}[!h]
    \centering
    \includegraphics[width=.41\textwidth, height=4.8cm]{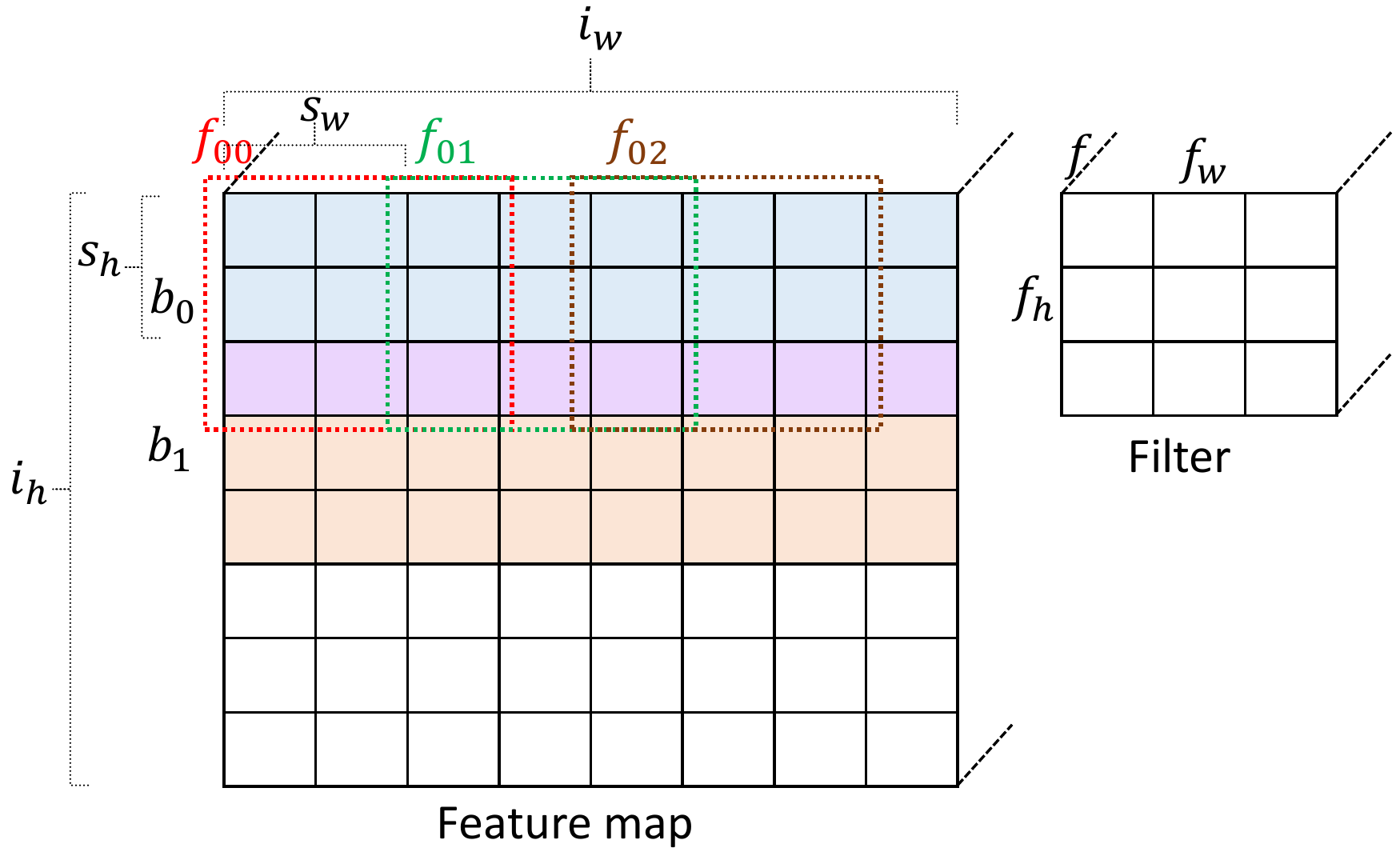}
    \caption{A batched processing of convolution and pooling layers for CIFAR-10 HCNN. Without loss of generality, the filter dimensions shown are set to $f_w = 3$ and $f_h = 3$, and the stride size is set to $s_w = 2$ and $s_h = 2$.}
    \label{fig:image:blocking}
\end{figure}

We first estimate the ciphertext size and calculate the maximum number of ciphertexts ($\eta$) GPU memory may contain. We horizontally partition the feature map into blocks, where the height of the block is equal to the filter height as shown in Figure~\ref{fig:image:blocking}. If the block size (in number of ciphertexts) is smaller than $\eta$, the entire block is copied to GPU memory, otherwise, we partition the block until its size is less than $\eta$. Note that the block size is a multiple of the filter size $f_w \cdot f_h$ to ensure a whole number of convolutions can be computed. Figure~\ref{fig:image:blocking} illustrates our mechanism to partition an input image (or feature map) with dimensions $i_w$ and $i_h$ into a number of blocks depending on the the filter dimensions $f_w$ and $f_h$ and stride sizes $s_w$ and $s_h$. The figure also illustrates how filter $f$ is scanned over the feature map to perform the convolution operation. The starting index of each filter placement $f_{jk}$ can be calculated as $f_{jk} = j\cdot s_h\cdot i_w + k\cdot s_w$ and the starting index of each block can be calculated as $b_j = j\cdot s_h \cdot i_w$. Note that performing the convolution over each block is completely independent and can be computed in parallel. On multiple GPU platforms, we ensure that each block is handled by a single CPU thread, which offloads computation to a specific GPU card.
\section{Experiments and Comparisons}\label{section:experiments}

\subsection{Hardware Configuration}
Table~\ref{tab:testBedConfiguration} shows the configuration of the testbed server used for all experiments. Note that on our system, the GPU cluster is heterogeneous and consists of 3 P100 cards and 1 V100. All single GPU experiments were invoked on the V100 card.
\begin{table}[!ht]
    \scriptsize
    \caption {Hardware configuration of the testbed servers} \label{tab:testBedConfiguration}
    \centering
    \begin{tabular}{@{}l l| ll@{}}
        \toprule
        {\bfseries Feature}    & \multicolumn{1}{c}{\bfseries CPU} & \multicolumn{2}{c}{\bfseries GPU Cluster}
        \\
        \midrule        
        {Model}               & Intel Xeon Platinum       & V100     & P100 \\
        {Compute Capability} & \multicolumn{1}{c|}{$-$}  & 7.0    & 6.0 \\
        {\# of processing units}        & 2                           & 1     & 3 \\ 
        {\# Cores (total)} & 26                          & 5120  & 3584\\
        {Core Frequency}   & 2.10 GHz                    & 1.380 GHz &  1.328 GHz \\ 
        {Memory Type}      & 72-bit DDR4                 & 4k-bit HBM2 &  4k-bit HBM2 \\
        {Memory Bandwidth} & 30 GB/sec                   & 732 GB/sec &  900 GB/sec \\
        {Memory Capacity}  & 187.5 GB                    & $16$ GB    & $3\times 16$ GB \\
        \midrule
        {PCI-e Bandwidth}  & \multicolumn{2}{c}{16 GB/sec} \\
        \bottomrule
    \end{tabular}
\end{table}
\subsection{Datasets}
\noindent
\textbf{MNIST.} The MNIST dataset~\cite{MNIST} consists of 60,000 images (50,000 in training dataset and 10,000 in testing dataset) of hand-written digits, each is a \(28\times28\) array of values between \(0\) and \(255\), corresponding to the gray level of a pixel.\\

\noindent
\textbf{CIFAR-10.} The CIFAR-10 dataset~\cite{krizhevsky2009learning} consists of 60,000 colour images (50,000 in training dataset and 10,000 in testing dataset) of 10 different classes. Each images consists of 32 $\times$ 32 $\times$ 3 pixels of values between \(0\) and \(255\).

\subsection{Micro-Benchmarks}
Our HCNNs use 6 FHE primitives: 1) Key generation (\textsf{KeyGen}), 2) Encryption (\textsf{Enc}), 3) Decryption (\textsf{Dec}), 4) Homomorphic addition (\textsf{HAdd}), 5) Homomorphic squaring (\textsf{HSquare}) and 6) Homomorphic multiplication of ciphertext by plaintext (\textsf{HMultPlain}). Table~\ref{table:benchmarks} shows these primitives and their latency in milliseconds using SEAL and \afv~on CPU and GPU, respectively. It can be clearly seen that \afv~outperforms SEAL by at least one order of magnitude. On average, \afv~provides 22.36$\times$, 18.67$\times$, 91.88$\times$, 4.40$\times$, 48.07$\times$, 334.56$\times$ and 54.59$\times$ speedup for \textsf{KeyGen}, \textsf{Enc}, \textsf{Dec}, \textsf{HAdd}, \textsf{HSquare}, \textsf{HMultPlain} and \textsf{HMult}.

The results show that \textsf{HSquare}, which is used in the activation layers, is the most time-consuming operation in our HCNNs. In contrast, both \textsf{HAdd} and \textsf{HMultPlain}, which are used in the convolution and fully connected layers, are very cheap. Note that our HCNNs can be modified to run an encrypted model on encrypted data. This can be done by replacing \textsf{HMultPlain} by \textsf{HMul}. However, this will affect the performance severely as \textsf{HMult} is the most expensive primitive in FHE. In addition, the storage requirement for encrypted models will be very huge. In fact, scenarios where the model has to be encrypted are not practical with FHE since each client requires a copy of the model encrypted with her own secret key.

\begin{table}[!ht]
  \scriptsize
  \centering
  \caption{FHE primitives benchmarks using SEAL and \afv~on CPU and GPU, respectively. Time unit is millisecond. Note that HMult was not used in HCNN.}
        \begin{tabular}{ccrrr}
    \toprule
    \multicolumn{1}{c}{\textbf{Function}} &  \begin{tabular}{@{}c@{}}\textbf{Parameter} \\ \textbf{ID}\end{tabular}  & \begin{tabular}{@{}c@{}}\textbf{SEAL} \\ \textbf{CPU}\end{tabular} & \begin{tabular}{@{}c@{}}\textbf{\afv} \\ \textbf{GPU}\end{tabular} & \multicolumn{1}{c}{\textbf{Speedup}} \\
    \midrule
    \multirow{2}[1]{*}{\textsf{KeyGen}} 
          & 2     & 272.142 & \textbf{12.377} & 21.99$\times$ \\
          & 4     & 542.920 & \textbf{21.392} & 25.38$\times$ \\
          \midrule
    \multirow{2}[0]{*}{\textsf{Enc}} 
          & 2     & 12.858 & \textbf{0.935} & 13.75$\times$ \\
          & 4     & 25.991 & \textbf{1.496} & 17.37$\times$ \\
          \midrule
    \multirow{2}[0]{*}{\textsf{Dec}} 
          & 2     & 5.171 & \textbf{0.075} & 68.95$\times$ \\
          & 4     & 10.408 & \textbf{0.098} & 106.20$\times$ \\
          \midrule
    \multirow{2}[0]{*}{\textsf{HAdd}} 
          & 2     & 0.126 & \textbf{0.052} & 2.42$\times$ \\
          & 4     & 0.281 & \textbf{0.054} & 5.20$\times$ \\
          \midrule
    \multirow{2}[0]{*}{\textsf{HSquare}} 
          & 2     & 69.588 & \textbf{1.679} & 41.45$\times$ \\
          & 4     & 138.199 & \textbf{2.371} & 58.29$\times$ \\
          \midrule
    \multirow{2}[0]{*}{\textsf{HMultPlain}} 
          & 2     & 7.680 & \textbf{0.033} & 232.73$\times$ \\
          & 4     & 15.694 & \textbf{0.035} & 448.40$\times$ \\
          \midrule
    \multirow{2}[1]{*}{\textsf{HMult}$^*$} 
          & 2     & 86.270 & \textbf{2.014} & 42.84$\times$ \\
          & 4     & 173.167 & \textbf{2.769} & 62.54$\times$ \\
    \bottomrule
    \end{tabular}%
  \label{table:benchmarks}%
\end{table}%
\subsection{HCNNs Performance}
Table~\ref{table:benchmarking:results} shows the runtime of evaluating our MNIST and CIFAR-10 HCNNs. As mentioned previously, we did not run CIFAR-10 with SEAL as it will take a huge latency and resources. It can be seen that \afv~outperforms SEAL significantly for all parameter sets. In particular, the speedup factors achieved are 138.56$\times$ under parameter set 2 (at 76-bit security level) and 258.92$\times$ under parameter set 4 (at 159-bit security level). The amortized time represents the per-image inference time calculated as the ratio between network evaluation latency and the number of packed images in a ciphertext. As the number of slots in ciphertext is equal to the ring dimension $N$, using parameter sets (3 and 4 where $N = 2^{14}$) we can classify the entire testing dataset of MNIST (10,000 images) in a single network evaluation since we can pack a certain pixel from all images in one ciphertext (see Figure~\ref{fig:mnist:packing}). On the other hand, with parameter sets (1 and 2) we can classify only 8192 images in a single network evaluation time. 
We include the timing of all parameter sets to give the reader an insight into the relation between FHE parameters and performance. On CPU, doubling $N$ results in almost doubling the inference latency (739.90 sec when $N=2^{13}$ vs. 1,563.85 sec when $N=2^{14}$), whereas on GPU, a much lower factor is noticed due to parallel execution. We note it is not recommended to use parameter set 2 as it only provides 76 bit security level. Note that SEAL fails to evaluate MNIST HCNN under parameter sets 1 and 3 due to higher noise growth in their underlying scheme. This will be discussed further in Section~\ref{subsec:noise:growth}.
\begin{table}[!ht]
    \centering
    \scriptsize
    \caption{Latency (in seconds) of running HCNNs with SEAL and \afv~on multi-core CPU and GPU, respectively. PIT refers to per-image time if packing is used.}
    \begingroup
\setlength{\tabcolsep}{4pt} 
\renewcommand{\arraystretch}{1} 
    \begin{tabular}{lcrrrrr}
        \toprule
        {HCNN} & {Param} & \multicolumn{2}{c}{{CPU}} & \multicolumn{2}{c}{{GPU}} & \multicolumn{1}{c}{{Speedup}} \\
        
        \cmidrule[0.4pt](r{0.125em}){3-4}
        \cmidrule[0.4pt](r{0.125em}){5-6}
        
         & ID & \multicolumn{1}{c}{SEAL} & \multicolumn{1}{c}{PIT$\times 10^{-3}$} & \multicolumn{1}{c}{\afv} & \multicolumn{1}{c}{PIT$\times 10^{-3}$} &  \\
        \midrule
        \multirow{4}{*}{MNIST$_{1\text{G}}$} & 1 & Failure & \multicolumn{1}{c}{$-$} & 5.16 & 0.63& \multicolumn{1}{c}{$-$}\\
        
        & 2 & 739.90 & 90.32 & 5.34 & 0.65 & 138.56$\times$ \\
        
        & 3 & Failure & \multicolumn{1}{c}{$-$} & 5.71 & 0.34 & \multicolumn{1}{c}{$-$} \\
        
        & 4 & 1,563 & 156.38 & 6.04 & 0.37 & 258.92$\times$\\[.1cm]

        CIFAR-10$_{1\text{G}}$ & 5 & \multicolumn{1}{c}{$-$} & \multicolumn{1}{c}{$-$} & 553.89 & $67.61$ & \multicolumn{1}{c}{$-$}\\
        
        CIFAR-10$_{4\text{G}}$& 5 & \multicolumn{1}{c}{$-$} & \multicolumn{1}{c}{$-$} & 304.43 & $37.16$ & \multicolumn{1}{c}{$-$}\\
        
        \bottomrule
    \end{tabular}%
    \endgroup
    \label{table:benchmarking:results}%
\end{table}%

The table also includes the latency of running our CIFAR-10 HCNN using \afv. We show the results of running our CIFAR-10 HCNN on 1 and 4 GPU cards. The latency shown here is per 1 plaintext modulus prime, i.e., 1 CRT channel. Note that we use 10 primes to evaluate CIFAR-10 HCNN. As our HCNNs will typically be hosted by the cloud, one may assume that 10 machines can evaluate CIFAR-10 HCNN in 304.430 seconds.

We also note that our timing results shown here for SEAL are much higher than those reported in CryptoNets (570 seconds at 80-bit security). This can be attributed to employing the YASHE\(^{\prime}\) levelled FHE scheme in CryptoNets, which is known to be less computationally intensive compared to BFV that is currently implemented in SEAL~\cite{lepoint2014comparison}. \caesar{We remark that the YASHE\(^{\prime}\) scheme is susceptible to the subfield lattice attack proposed by Albrecht \etal~\cite{C:AlbBaiDuc16}.} 

Lastly, we compare our best results with the currently available solutions in the literature. Table~\ref{table:literature:comparison} shows the reported results of previous works that utilized FHE to evaluate HCNNs on different datasets. As we can see, our solution outperforms all solutions in total and amortized time. For instance, our MNIST HCNN is 110.47$\times$, 5.54$\times$ and 6.85$\times$ faster than CryptoNets, E2DM and Faster CryptoNets (FCryptoNets), respectively. Note that E2DM classifies 64 images in a single evaluation. Similarly, our CIFAR-10 HCNN is 3.83$\times$ and 7.35$\times$ faster than CryptoDL and FCryptoNets, respectively when it is evaluated on a single machine. On 10 machines, one can get an extra 10$\times$ factor of speedup. We remark that batch size in these solutions depends on the packing mechanism used. For instance, it is a constant in CryptoNets and HCNN where it is equal to the ring dimension $N$. In FCryptoNets, the batch size is equal to 1 as no packing mechanism is used. E2DM uses a different packing mechanism where the entire image is packed in a single ciphertext, therefore, batch size is the ratio between $N$ and image size.

\caesar{
We remark that the reported runtimes in Table~\ref{table:literature:comparison} were observed on different platforms. For instance, CryptoNets was evaluated on a single Intel Xeon E5-1620 CPU running at 3.5 GHz, with 16 GB of RAM~\cite{MSFT:DGL+16}. E2DM was evaluated on a Macbook Pro laptop with an Intel Core i9 running with 4 cores rated at 2.3 GH ~\cite{Jiang:2018:SOM:3243734.3243837}. FCryptoNets evaluated their MNIST HCNN on a machine with Intel
Core i7-5930K CPU at 3.5 GHz with 48 GB RAM, whereas their CIFAR-10 HCNN was evaluated on n1-megamem-96 instances from Google Cloud Platform, which each has 96 Intel Skylake 2.0 GHz vCPUs and 1433.6 GB RAM~\cite{DBLP:journals/corr/abs-1811-09953}. Lastly, CryptoDL ran their experiments on a computer with 16 GB RAM and Intel Xeon E5-2640 CPU rated at 2.4 GHz.
}
        
        
        
        
        
        
        
\begin{table}[h]
    \centering
    \scriptsize
    \caption{Comparison of runtime (seconds), security level and accuracy between prior FHE-based HCNNs and our HCNNs.}
        \begingroup
\setlength{\tabcolsep}{4pt} 
\renewcommand{\arraystretch}{1} 
    \begin{tabular}{p{2cm}rrrrc}
        \toprule
        {Model} & \multicolumn{2}{c}{{Runtime (sec)}} & $\lambda$ & Accuracy (\%) & Dataset\\
        
        \cmidrule[0.4pt](r{0.125em}){2-3}
        
         & \multicolumn{1}{c}{Total} & \multicolumn{1}{c}{Amortized time} \\
        \midrule
        {CryptoNets~\cite{MSFT:DGL+16}} & 570 & 69.580$\times 10^{-3}$ & 80 & 99.00 & MNIST\\
        
        {E2DM~\cite{Jiang:2018:SOM:3243734.3243837}} & 28.590 & 450.0$\times 10^{-3}$ & 80 & 98.01 & MNIST\\
        
        \multirow{1}{2cm}{{FCryptoNets~\cite{DBLP:journals/corr/abs-1811-09953}}} & \multirow{1}{*}{39.100} & \multirow{1}{*}{39.100} & \multirow{1}{*}{128} & 98.71 & \multirow{1}{*}{MNIST}\\
        
        {\afv} & 5.160 & 0.630$\times 10^{-3}$ & 82 & 99.00 & MNIST\\
        {\afv} & 5.710 & 0.340$\times 10^{-3}$ & 175& 99.00 & MNIST\\
        
        \midrule
        
        {CryptoDL~\cite{DBLP:journals/corr/abs-1711-05189}} & 11,686 & 11,686 & 80 & 91.50 & CIFAR-10\\
        
        \multirow{1}{1.5cm}{{FCryptoNets~\cite{DBLP:journals/corr/abs-1811-09953}}} &  \multirow{1}{*}{22,372} & \multirow{1}{*}{22,372} & \multirow{1}{*}{128} & 75.99 & \multirow{1}{*}{CIFAR-10}\\
        {\afv} & 304.43 & 0.372 & 91 & 77.55 & CIFAR-10\\
        \bottomrule
    \end{tabular}%
    \endgroup
    \label{table:literature:comparison}%
\end{table}%

\subsection{Noise Growth}\label{subsec:noise:growth}
\begin{figure*}[h]
     \centering
     \subfloat[][Parameter Set 3]{\includegraphics[width=0.42\linewidth,height=4cm]{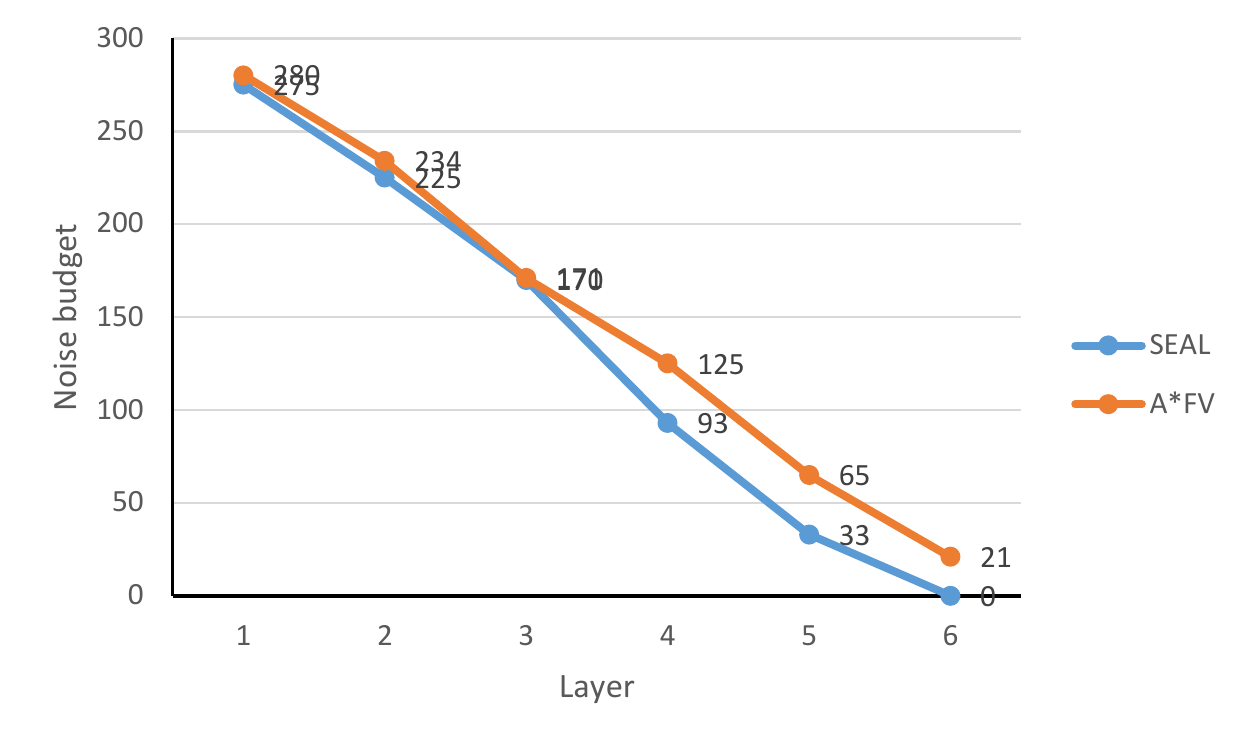}\label{fig:noise:budget:paramset3}}
     \subfloat[][Parameter Set 4]{\includegraphics[width=0.42\linewidth,height=4cm]{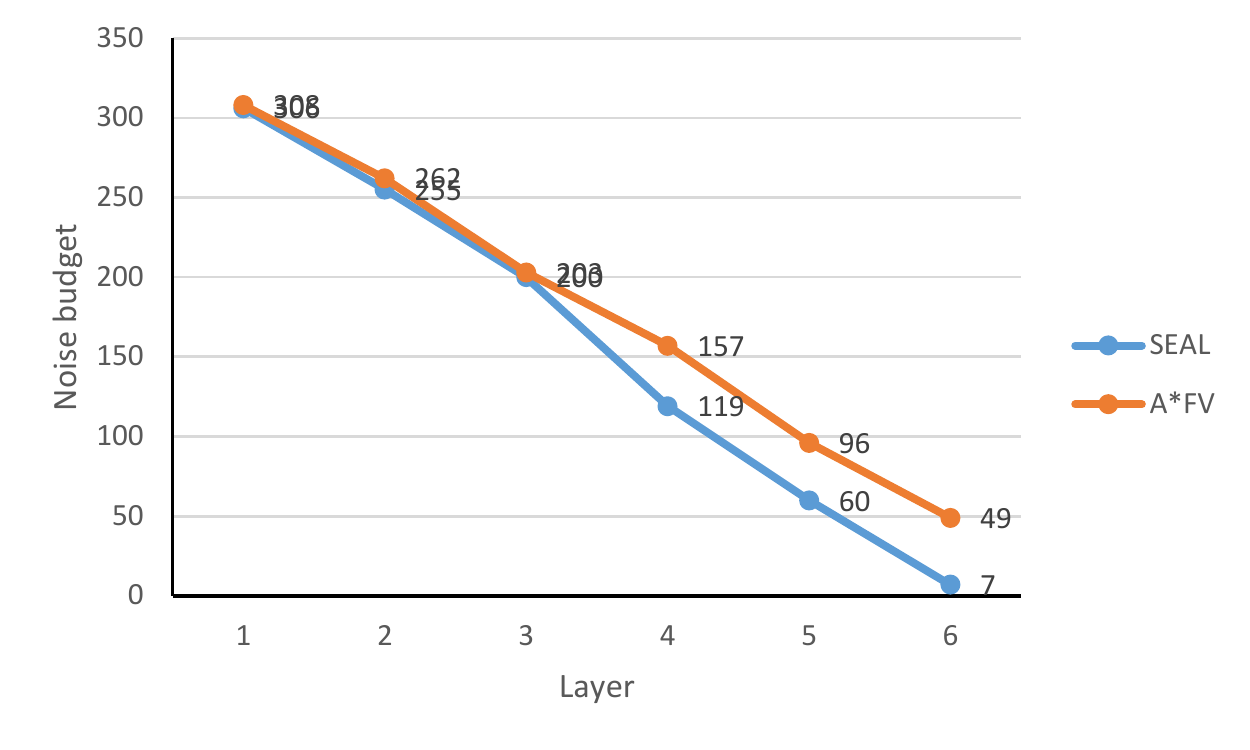}\label{fig:noise:budget:paramset4}}
     \caption{Noise budget (in bits) and how it is consumed by SEAL and \afv~}
     \label{fig:noise:budget}
\end{figure*}
In this section, we show the noise growth behaviour in both SEAL and \afv. We recall that SEAL version (2.3.1) implements the BEHZ~\cite{bajard2016full} RNS variant of the BFV scheme. On the other hand, \afv~implements a different RNS variant known as the HPS~\cite{EPRINT:HalPolSho18}. Although both variants implement the same scheme, it was found by Al~Badawi~\etal~\cite{EPRINT:BPAVR18} that these variants exhibit different noise growth behaviour. Figure~\ref{fig:noise:budget} shows the noise growth behaviour in both SEAL and \afv~for the parameter sets 3 and 4. The vertical axis represents the noise budget which can be viewed as the ``signal-to-noise'' ratio. Once the noise budget reaches 0 in a ciphertext, it becomes too noisy to compute further or to decrypt successfully. As seen in the figure, parameter set 3 is not sufficient to provide SEAL with sufficient noise budget to evaluate the MNIST HCNN. The ciphertexts generated by the fully connected layer include noise budget 0. Although no further computation is required after the fully connected layer, decryption fails due to the high noise level. On the other hand, \afv~has lower noise budget consumption rate that is sufficient to run MNIST HCNN with some left out noise budget (21 bits in parameter set 3 and 49 bits in parameter set 4). Parameter set 4 provides higher noise budget that is sufficient to run the MNIST HCNN in both SEAL and \afv~successfully. 

\section{Conclusion}\label{section:conclusion}
In this work, we presented a fully FHE-based CNN that is able to homomorphically classify encrypted images. The main motivation of this work was to show that privacy-preserving DL with FHE is dramatically accelerated with GPUs and offers a way towards efficient DLaaS. Our implementation included a set of techniques such as low-precision training, unified training and testing network, optimized FHE parameters and a very efficient GPU implementation to achieve high performance. We evaluated our system on two different datasets MNIST and CIFAR-10. Our solution achieved high security level ($> 80$ bit) and reasonable accuracy (99\%) for MNIST and (77.55\%) for CIFAR-10. In terms of performance, our best results show that we could classify a batch of images (8192) in 5.16 and 304.43 seconds for MNIST and CIFAR-10, respectively. 



\section*{Acknowledgment}
\caesar{This work is supported by A*STAR under its RIE2020 Advanced Manufacturing and Engineering (AME) Programmtic Programme  (Award A19E3b0099). Any opinions, findings and conclusions or recommendations expressed in this material are those of the authors and do not reflect the views of A*STAR.}

\bibliographystyle{unsrt}
\bibliography{biblio}

\vfill
\end{document}